%% file: main.tex
\documentclass[10pt,journal,compsoc]{IEEEtran}
\input{files/packages}
\input{files/acronyms}
\begin{document}
\input{files/title_authorship}

\IEEEtitleabstractindextext{
\begin{abstract}
Shared L1-memory clusters of streamlined instruction processors (processing elements - \glspl{pe}) are commonly used as building blocks in modern, massively parallel computing architectures (e.g. GP-GPUs).
\emph{Scaling out} these architectures by increasing the number of clusters incurs computational and power overhead, caused by the requirement to split and merge large data structures in chunks and move chunks across memory hierarchies via the high-latency global interconnect.
\emph{Scaling up} the cluster reduces buffering, copy, and synchronization overheads.
However, the complexity of a fully connected cores-to-L1-memory crossbar grows quadratically with \gls{pe}-count, posing a major physical implementation challenge.
We present \terapool{}, a physically implementable, $>$\num{1000} floating-point-capable RISC-V \glspl{pe} scaled-up cluster design, sharing a Multi-MegaByte $>$\num{4000}-banked L1 memory via a low latency hierarchical interconnect (\num{1}-\num{7}/\num{9}/\num{11} cycles, depending on target frequency).
Implemented in \SI{12}{\nano\meter} FinFET technology, \terapool{} achieves near-gigahertz frequencies (\num{910}\si{\mega\hertz}) typical, \SI{0.80}{\volt}/\SI{25}{\celsius}.
The energy-efficient hierarchical \gls{pe}-to-L1-memory interconnect consumes only \num{9}-\SI{13.5}{\pico\joule} for memory bank accesses, just \num{0.74}-\SI{1.1}{\times} the cost of a FP32 FMA.
A high-bandwidth main memory link is designed to manage data transfers in/out of the shared L1, sustaining transfers at the full bandwidth of an HBM2E main memory.
At \SI{910}{\mega\hertz}, the cluster delivers up to \num{1.89} single precision \si{\tera\flop\per\second} peak performance and up to \SI{200}{\giga\flop\per\second\per\watt} energy efficiency (at a high \si{IPC\per\gls{pe}} of \num{0.8} on average) in benchmark kernels, demonstrating the feasibility of scaling a shared-L1 cluster to a thousand \glspl{pe}, four times the \gls{pe} count of the largest clusters reported in literature.
\end{abstract}

}

\maketitle
\glsresetall
\copyrightnotice

\section{Introduction}
\label{sec:introduction}
\IEEEPARstart{T}{h}e computing requirements of Generative \gls{ai}, digital twins, 6G wireless networks, autonomous vehicles, and robots doubles every \num{6}-\num{9} months~\cite{Compute_Trend_2022}.
This exponential growth escalates to quadrillions of \si{FLOPs} and requires substantial memory capacity to manage massive high-speed data streams and \gls{ml} models with trillions of parameters~\cite{LLM_Survey_2_2024, Survey_Hardware_Acc_2022}.
The success in fast-paced adaptation, scale-up, and evolution of these AI-centric, data-intensive workloads relies on highly programmable high-performance hardware, with streamlined programming models and toolchains.
Steadily increasing performance demands under tight power bounds, while maintaining a high degree of programming flexibility and ease-of-use, steers computing architectures toward flexible, but scalable and energy-efficient solutions~\cite{50Years_trend_2022, Near_Mem_2019}.

\begin{figure}[htbp]
  \centering
  \includegraphics[width=\linewidth]{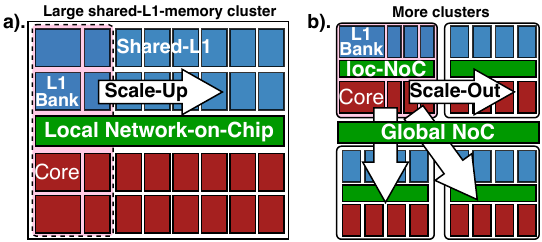}
  \caption{Illustration of scaling-up and scaling-out in cluster(s) design.}
  \label{fig:cluster_scaling}
\end{figure}

In this challenging context, many-core architectures with a large number of streamlined programmable \glspl{pe} running at moderate (\num{1}-\SI{2}{\giga\hertz}) frequency, have achieved much higher efficiency than architectures based on a few large processors running at extremely high frequency (\num{5}$+$\si{\giga\hertz})~\cite{Energy_Efficient_2022,Energy_problem_2014}.
The most common architectural building block in modern many-core designs is the cluster of \glspl{pe} tightly coupled with shared-L1 memory through a low-latency, high-bandwidth interconnect.
However, the cluster is typically limited in scale to tens to low hundreds of \glspl{pe}~\cite{mppa256_2017, ET_soc_2022, Ramon_2021}.
Beyond this number of \glspl{pe}, performance is usually increased by scaling out (see~\cref{fig:cluster_scaling}b) to many loosely-coupled clusters, which access the main memory and other clusters' L1 memory through a high-latency global interconnect.
However, loosely-coupled clusters imply hardware and software overheads, namely: synchronization, inter-cluster communication, data allocation-splitting, and workload distribution~\cite{Multi_Million_Core_2021}. 
To contain these performance losses and enhance energy efficiency, directly scaling up (see~\cref{fig:cluster_scaling}a) the shared-L1 cluster is highly desirable~\cite{Towards_Energy_Eff_Cluster_2012}.
For instance, \emph{NVIDIA}'s \gls{gpu} architecture, which follows the shared-L1 architectural model, features clusters called \glspl{sm} with multi-threaded tensor cores sharing L1 data memories.
From the \emph{A100}~\cite{nvidia_a100_2020} to the \emph{H100}~\cite{nvidia_h100_2023}, the peak per-\gls{sm} \gls{fp} computational power scaled up $4\times$ by doubling the number of \gls{fp}-\glspl{pe} with increased processing rates.
Additionally, the shared memory size expanded from \SI{192}{\kilo\byte} to \SI{256}{\kilo\byte}.

The \gls{pe} to L1 memory interconnect design emerges as the main challenge to cluster scale-up from both the architectural and physical design viewpoints:
\viewpoint{1} The complexity of a baseline crossbar interconnect between \glspl{pe} and memory banks would scale quadratically with the number of \glspl{pe} and memory banks in the cluster~\cite{interconnect_2020}, becoming a blocking factor in physical implementation.
\viewpoint{2} Keeping low latency memory access in large-scale, unified-address, and physically distributed banks is essential for minimizing \gls{pe} stalls and ensuring high computing utilization.
\viewpoint{3} In addition, the requirement also provides a link between the L1 memory within a cluster and the rest of the memory hierarchy further straining the physical routing resource budget.
Therefore, preserving the cluster's \gls{ppa} while scaling up requires carefully balancing the physical feasibility and interconnect performance.
These interconnect design challenges have so far limited cluster scale-up to tens or, at most, hundreds of \glspl{pe}~\cite{Mempool_2023}.

We present \terapool{}, a physically feasible, energy-efficient, general-purpose, and programming-friendly many-core cluster design.
We break the current scale-up barrier, achieving 1000-\glspl{pe} shared-L1 cluster, while maintaining near-GHz frequencies and delivering \si{\tera\flop\per\second} performance.
Extending our earlier work~\cite{TeraPool_2024}, we presents the following contributions:
\begin{itemize}[leftmargin=*]
    \item A large scale shared-L1 cluster design, based on a hierarchical crossbar interconnect, \num{4} times larger than the largest reported in the literature~\cite{Mempool_2023}, featuring \num{1024} \glspl{pe} (RV32IMAF cores) sharing \SI{4}{\mebi\byte} (\num{4096} banks) of L1 \gls{spm};
    \item An \gls{amat} model of the hierarchical crossbar interconnect between \glspl{pe} and memory banks. The model's insight drives the design and optimization of a three-level hierarchical \gls{pe}-L1 interconnect (\emph{Tile}, \emph{SubGroup}, and \emph{Group}), achieving \gls{numa} latencies of \num{1}-\num{5} cycles within a Group and up to \num{7}-\num{11} cycles for remote Group access, depending on the target operating frequency;
    \item A \gls{hbml} supporting data movement in and out of the cluster's L1 banks with negligible overhead. Co-simulated with an open-source, cycle-accurate \emph{DRAMsys5.0} memory model~\cite{DRAMsys_2022} with hybrid address mapping scheme, our \gls{hbml} achieves upto \SI{97}{\percent} bandwidth utilization of an industry-standard \gls{hbm};
    \item The performance evaluation across key data-parallel kernels, demonstrates \terapool{}'s capability on large data chunks with minimal data-transfer overhead.
    \item A detailed physical design methodology, floorplan and \gls{ppa} analysis in GlobalFoundries' \SI{12}{\nm} FinFET technology.
\end{itemize}

The remainder of this paper is structured as follows:
\cref{sec:motivation} details the motivation for scaling up shared-L1 cluster.
\cref{sec:interconnection} introduces the \gls{amat} modeling for hierarchical \gls{pe}-to-L1 interconnect, evaluating the design choices through latency, throughput, and physical routing complexity.
\cref{sec:cluster} describes the core-complex design and hierarchical cluster architecture.
Next, \cref{sec:system} discusses \gls{hbml} design, the \emph{DRAMsys5.0} co-simulation, and the hybrid memory address mapping scheme.
The physical implementation and \gls{ppa} analysis is detailed in~\cref{sec:physical}.
In~\cref{sec:software}, we present the \terapool{}'s streamlined programming model and software performance.
Finally, \cref{sec:soa} and~\cref{sec:conclusion} discuss related work and conclusions, respectively.
Our design, including open \emph{DRAMsys5.0} model support and the full software infrastructure, is open-sourced under a liberal license~\footnote{\text{https://github.com/pulp-platform/mempool}}.

\section{Shared-memory cluster scale-up vs. scale-out}
\label{sec:motivation}
In this work, \emph{Cluster} refers to a computing block where \glspl{pe} share L1-memory through a low-latency interconnect. Its main design parameters are summarized in~\Cref{tab:cluster_parameter}.
This section discusses the scaling strategies, highlighting the benefits and trade-offs from: \viewpoint{1} scaling up in a single cluster, and \viewpoint{2} scaling out to multiple clusters.

\begin{table}[!ht]
\centering
\caption{Cluster Design Parameters Description}
\begin{tabularx}{\linewidth}{rX}
\toprule
\textbf{Symbol} & \textbf{Description}                                                  \\\midrule
$L$             & Main memory latency (cycle)                                           \\
$W$             & Problem-tiling size in L1 memory (word)                               \\
$BW$            & Bandwidth between cluster\&main memory  (words/cycle)                 \\
$AI$             & Arithmetic intensity (operations/word)                               \\
$N_{PEs}$       & Number of processing elements in a cluster                            \\
$U$             & Computing utilization of a processing element                         \\
$S$             & Scaling factor for PEs and memory banks in the cluster                \\\bottomrule
\end{tabularx}
\label{tab:cluster_parameter}
\end{table}

\subsection{Single Cluster Scale-up}
\label{sec:scaling-up}

Since the problem size typically exceeds the cluster’s L1 memory capacity, tiling is commonly adopted: large data structures are split into chunks and moved in L1.
Scaling up the cluster increases the number of processing elements and the total shared-L1 capacity for larger problem tiling, while the larger die area accommodates more I/Os, enabling higher $BW$ between L1 and main memory.
We analyze two workload cases to show the scale-up benefits:

\noindent\circled{1} For workloads \textbf{without} data reuse, the $AI \leq 1$ typically leads to performance being main-memory bound.
Small clusters may need extra inter-cluster data transfers.
Additionally, larger data chunks improve workload balance across \glspl{pe}, reduce synchronization overhead, and increase resilience to main memory latency and bandwidth jitter.

\noindent\circled{2} For workloads \textbf{with} data reuse, a larger L1 capacity reduces cluster to main-memory $BW$ demands.
Consider an example \gls{matmul}, where each matrix chunk has dimension $m$, $W=3m^2$ and $AI={m^3}/{3m^2}={\sqrt{W}}/{3\sqrt{3}}$. 
During cluster scaling by a factor of $S$, $W$ scales linearly, leading to an increase in $AI$ that reflects greater data reuse, given by:
\begin{align}
    & W' = S \times W, AI' = \frac{\sqrt{SW}}{3\sqrt{3}} = \sqrt{S} \times AI
\label{eq:motivation_1}
\end{align}
Following the \emph{Kung's Principle}~\cite{Kung_1986}, the cluster is not bottlenecked by main memory bandwidth and latency when~\Cref{eq:motivation_2} holds:
considering cluster scaling by a factor $S$, we assume $L$ and $U$ remain constant due to identical design elements.
The left-hand side represents the data transfer costs, where both $W$ and $BW$ scale with $S$ and their ratio remains unchanged.
The right-hand side reflects computational demand; the ratio of $N_{\mathrm{PEs}}$ to $W$ remains unchanged as both scale with $S$, while $AI$ increases with $\sqrt{S}$, improving the compute-to-memory balance.
\begin{equation}
    L + \frac{W}{BW} < \frac{\sqrt{S}AI \times W}{N_{PEs} \times U}
\label{eq:motivation_2}
\end{equation}

This reveals that as the scaling factor $S$ increases, the inequation holds for larger $L$ and smaller $BW$, demonstrating that large-scale shared-L1-memory clusters are better equipped to maintain high computational utilization while tolerating main memory large transfer latency and limited bandwidth.

\subsection{Scale-out to Multiple Clusters}
\label{sec:scaling-out}
As an alternative to the scale-up strategy, several smaller clusters can be instantiated, each accessing main memory and other clusters through global interconnects, representing a loosely coupled scale-out architecture~(\cref{fig:cluster_scaling}b).
Compared with equivalent computing and memory resources in cluster scaling up, scaling out to a multi/many-cluster system costs computational and power overheads, leading to underutilized hardware components and reduced energy efficiency~\cite{Scale_up_out_2016}.
We identify three common overheads that increase with the scaling factor:
\begin{itemize}[leftmargin=*]
    \item \textbf{Synchronization Overhead:} The processing times across clusters are hard to perfectly balance due to independent instruction flows, L1-memory refill ordering, and main memory access conflicts. This uneven utilization between clusters leads to synchronization overhead that increases with the number of clusters due to long tail effects~\cite{tail_scale_2013}.
    \item \textbf{Tiling Overhead:} Problem-tiling across loosely coupled clusters introduces additional computational overhead during data structure splitting and partial results merging across memory hierarchies. For instance, the \emph{Model Parallel}~\cite{Illusion_2021} adopted in \glspl{dnn} parallelization requires partial result reduction sums, where the execution time increases with the cluster counts.
    \item \textbf{Data Transfer Overhead:} Problem-tiling often leads to data duplication, as shared data must be copied to each cluster. Additionally, partial results from each cluster must be merged by transferring back to the main memory and then redistributed for subsequent processing through a high-latency global interconnect. These factors reduce the clusters' memory utilization and introduce data transfer overheads that grow with the number of clusters~\cite{TeraPool_2024}.
\end{itemize}
Additionally, connecting multiple clusters to main memory requires a high-latency global \gls{noc}, incurring additional hardware overhead.
A substantial die area is devoted to inter-cluster \gls{noc} routing, which consequently reduces area utilization and power efficiency~\cite{Occamy_2025}.

These considerations motivate our exploration of efficient cluster scale-up, aiming to achieve high compute utilization with physically feasible interconnects.

\section{interconnect Design}
\label{sec:interconnection}
Bandwidth and latency are the key performance metrics in the design of the \glspl{pe}-to-L1 interconnect.
Maintaining low latency ensures that \gls{pe} can hide it by issuing a low number of non-blocking L1 memory accesses, thereby preventing \gls{lsu} stalls and maintaining high utilization.
Although high diameter \gls{noc} topologies with low-cardinality switches can achieve high clock speed, they are not optimal for the \glspl{pe} to L1-memory interconnect, because latency increases by a few cycles per hop.
This leads to \gls{pe} throughput degradation, because too many outstanding memory transactions need to be supported to hide the \gls{noc} latency~\cite{cop_2023}.
Ideally, the shortest latency is achieved by a fully-combinational and \gls{fc} logarithmic-staged crossbar~\footnote{A logarithmic-staged tree topology; routing/arbitration implemented by demultiplexers/multiplexers with combinational logic control provides fine-grained address interleaving and single-cycle latency.}~\cite{interconnect_2020}, providing single-cycle data transfer between the \glspl{pe} and \gls{spm} banks.
The handshaking protocol has forward requests including the address, data write, and control signals, and backward responses containing the request ID, read data, and acknowledgment signals.
Simultaneous requests on the same arbitration switch are handled by a round-robin strategy.

\begin{figure}[!ht]
  \centering
  \includegraphics[width=\linewidth]{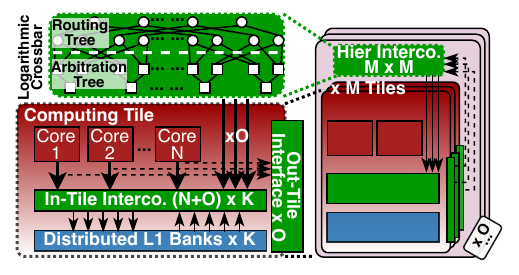}
  \caption{The hierarchical \emph{Tile} (building block) design with logarithmic-staged crossbar interconnect.}
  \label{fig:interco_diagram}
\end{figure}

Unfortunately, it has been shown~\cite{Cdxbar_2019} that while the crossbar is latency-optimal, its quadratic wire routing complexity poses insurmountable obstacles to achieving high frequencies.
Ultimately, excessively large crossbars become non-routable during physical design.
Hence, we need to find a compromise between latency, operating frequency, and physical routability.
We resort to a hierarchical design.
The hierarchical pattern builds upon basic building blocks, called \emph{Tiles} (in~\cref{fig:interco_diagram}), where a \gls{fc} crossbar links a physically neighboring group of \glspl{pe} to L1 \gls{spm} banks.
In higher hierarchy levels \emph{Tiles} are grouped, and \glspl{pe} access the L1 banks of other \emph{Tiles} via remote access ports. These ports connect to Tile-Tile crossbar(s) in the next hierarchy level, forming a fully shared-memory cluster.
Pipeline registers are introduced at hierarchy boundaries to mitigate long physical paths and improve operating frequency at the price of increased latency.
This pattern leads to a \gls{numa} architecture in which both the request and response paths feature complete arbitration.

In this section, we present the \gls{amat} model of the hierarchical \gls{pe}-to-L1 interconnect.
Furthermore, we leverage the model's insights and explore the physical design feasibility of various hierarchical design choices, demonstrating the optimal configuration for building an extremely scaled-up cluster with \num{1024} cores shared \num{4096} L1-\gls{spm} banks.

\subsection{Hierarchical Interconnect AMAT Model}
The \gls{amat} modeling analysis is conducted from the \emph{Tile} perspective where the memory requests are received, with the relevant parameters summarized in~\cref{tab:interco_parameter}.
Memory requests originate from either \emph{local-Tile} (requests sent from \glspl{pe} within the same Tile) or \emph{remote-Tile} (requests sent from \glspl{pe} in other Tiles).
Considering uniformly distributed random address requests, the probability that a request is sent from the \glspl{pe} within its local-Tile, denoted as $P_{Ltile}$, is given by ${1}/{N_{Tiles}}$.
The cluster's \gls{amat}, which contributes from both crossbar contention and pipeline registers, is computed as a probability-weighted sum of the latencies from each source, as shown in the equation below:
\begin{equation} 
    \begin{aligned}
        &T_{cluster} = P_{Ltile} \cdot L_{Ltile} + P_{Rtile} \cdot L_{Rtile}\\
        &P_{Rtile} = 1 - P_{Ltile} = 1 - \frac{1}{N_{Tiles}}\\
        &L_{L/Rtile} = E_{L:\ n\times k} + L_{pipeline}
\end{aligned}
\label{eq:interco_1} 
\end{equation}
Contention in the logarithmic-staged crossbar arises when multiple requests traverse the same arbitration tree node, targeting the same hierarchical port, thereby introducing an additional cycle of latency for each request due to arbitration.
This contention latency $E_{L:\ n\times k}$ can be modeled based on the probability of multiple requests passing through the same arbitrator node within a single cycle.

\begin{table}[!ht]
\centering
\caption{interconnect Modeling Parameter Descriptions}
\begin{tabularx}{\linewidth}{rX}
\toprule
\textbf{Symbol} & \textbf{Description}  \\\midrule
$N_{Tiles}$               & Number of building blocks (\emph{Tiles}) in a cluster \\
$T_{cluster}$             & Cluster's AMAT when all PEs perform random address access \\
$P_{Ltile/Rtile}$         & Probability that a memory request originates from the local/remote Tile  \\
$L_{Ltile/Rtile}$         & Memory access latency for requests sending from the local/remote Tile  \\
$L_{pipeline}$            & Latency introduced by pipeline registers \\
$n, k$                    & Number of crossbar inputs, outputs \\
$p$                       & Request injection rate at crossbar input \\
$P_{req}(x)$              & Probability that $x$ requests are sent to an arbitrator in a single cycle \\
$E_{L:\ n\times k}$       & Expected latency for random target access through an $n \times k$ crossbar \\
\bottomrule
\end{tabularx}
\label{tab:interco_parameter}
\end{table}

Our analysis begins with a simple arbitrator and extends to a multi-stage hierarchical crossbar design.
First, a basic \emph{N-to-1} arbitrator model forms the foundation of the contention analysis.
The arbitrator model is based on a Binomial($n$, $p$) distribution, where each input is considered an independent random trial, with $n$ representing the number of inputs and $p$ the probability of a request.
The \gls{pmf} $P_{req}(x)$ of binomial distribution represents the probability that $x$ requests are sent to the arbitrator within a cycle.
With $x$ initiators sending a request, the latency is $x - 1$.
The latency for no requests ($x=0$) is zero, as no access is present in the arbitrator.
To compute the expected latency, we sum the probabilities of having $x = 1 \rightarrow n$ requests per cycle:
\begin{equation}
    E_{L:\ n\times 1} = \sum_{x=1}^{n} L(x) P_{req}(x) = \sum_{x=1}^{n} (x - 1) P_{req}(x)
    \label{eq:interco_3}
\end{equation}
Second, we expand this analysis to an \emph{n-to-k} arbitrator model.
When a fully random address request is received, the probability of it being forwarded to watch-point output where directed to the observed Tile is $\frac{1}{k}$.
The number of requests arriving at the same output still follows a Binomial($n$, $p/k$) distribution with an injection rate of $p$, while the remaining output ports form an \emph{n-to-(k-1)} arbitrator.
Consequently, the expected latency must be calculated using a recursive method.
If no request arrives at the watch-point output ($x=0$), it is redirected to other outputs, contributing to an increase in the average latency, with a latency of $E_{L:\ n\times(k-1)}$.
The \gls{pmf} for the average contention-cost latency can be expressed as:
\begin{equation}
    E_{L:\ n\times k} = 
        \begin{cases} 
            E_{L:\ n\times1} + P_{req}(0) E_{L:\ n\times(k-1)}, & k > 1 \\
            \sum_{x=1}^{n} (x-1) P_{req}(x), & k = 1
        \end{cases}
\label{eq:interco_4}
\end{equation}

Finally, to analyze the multi-stage hierarchical interconnect crossbars, the probability of a request arriving at the input of each stage is determined by the probability that the request was forwarded from the output of the previous stage:
\begin{equation}
    p^{stage(N)} = P^{stage(N-1)}_{req} = 1 - P^{stage(N-1)}_{req}(0)
\label{eq:interco_5}
\end{equation}
To enhance the accuracy of the \gls{amat} calculation under conditions where multiple contentions remain unresolved within a single cycle, pending requests dynamically adjust the crossbar's injection rate.  
We introduce input queues at each hierarchical crossbar stage\footnote{Our modeling with input queues is performed using Python~\num{3.7} scripts.} to simulate request queuing for dynamic injection rate adjustments.

\subsection{TeraPool interconnect Analysis}
\label{sec:interco_analysis}
\begin{figure}[htbp]
  \centering
  \includegraphics[width=\linewidth]{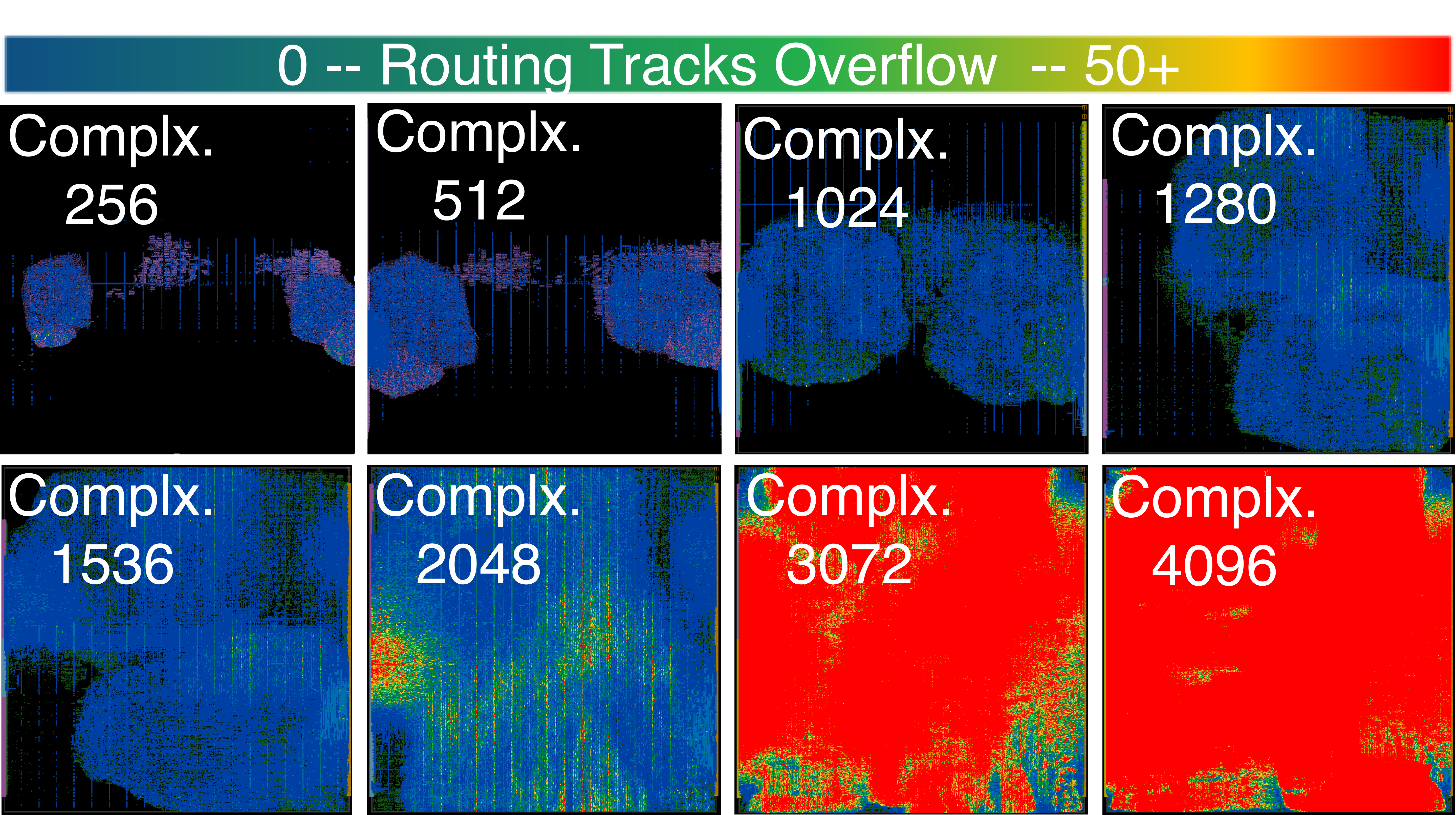}
  \caption{Routing Congestion of Logarithmic-Crossbar-Based Interconnect at Different Complexities (GF12nm, 13M).}
  \label{fig:interco_routing}
\end{figure}

In this subsection, we analyze various hierarchical multi-stage crossbar interconnect designs to determine the optimal \terapool{} architecture, extremely scaled-up to \num{1024} \glspl{pe} with \num{4096} shared-L1 banks. 
We evaluate the interconnect performance based on the following metrics:
\begin{itemize}[leftmargin=*]
    \item \textbf{Zero-load latency:} 
    The average latency cycles for a single random-address memory request without arbitration contention or input queuing. It is computed as the sum of each hierarchical access latency weighted by its access probability, reflecting the ideal latency of the interconnect.
    
    \item \textbf{AMAT:} 
    The average latency cycles taken for all \glspl{pe} to send random-address memory requests in the same cycle, factoring in arbitration contention and pipelining cycles, which are discussed in the previous subsection.

    \item \textbf{Throughput:} 
    The maximum requests injection rate through interconnect, expressed as \si{req/pe/cycle}. During continuous request injection, $L_{pipeline}$ can be hidden but throughput decreases caused by arbitration contention, as $1/E_{L:\ n \times k}$.

    \item \textbf{Interconnect complexity:} 
    The number of crossbar leaf nodes across internal routing and arbitration stages reflects routing complexity, and is approximated as $n \times k$.

    \item \textbf{Combinational delay:} 
    The number of routing levels. Requests traverse $\log_{2}n$ stages of the routing tree to reach one of the $n\times k$ leaf nodes, followed by $\log_{2}k$ levels of arbitration switches to the target memory bank. This delay is approximated as $\log_{2}n + \log_{2}k$.
\end{itemize}
Moreover, to assess the physical feasibility of the most complex stage in the hierarchical crossbar (referred to as \emph{Critical Complex.} in~\cref{tab:interconnection}), we evaluate routing congestion for interconnect complexity ($n\times k$) ranging from \num{256} to \num{4096}, using GlobalFoundries' \SI{12}{\nano\meter} FinFET technology with \num{13} metal stacks.
As shown in~\cref{fig:interco_routing} and~\cref{tab:congestion}, when complexity $<$\num{2048}, the interconnect remains routable. Each doubling in complexity increases the area and the wires by \num{1.8}$\times$, and the critical path delay by $<$\num{1.3}$\times$.
Beyond \num{2048}, routing becomes infeasible, resulting in \num{25}-\SI{308}{\percent} (\num{12.4}$\times$) \gls{beol} resource overflow, coupled with increased critical path delays that constrain cluster performance.

\begin{table}
    \caption{Routing Quality of Logarithmic-Staged Crossbar Interconnect at Different Complexities (GF12nm, 13M)}
    \label{tab:congestion}
    \centering
    \input{tables/congestion}
    \\ \footnotesize\raggedright\textsuperscript{*} The average routing track overflow rate for horizontal, vertical layers, and overall design.
\end{table}

\begin{table*}
    \caption{Hierarchical Interconnect Design Analysis for \num{1024} \glspl{pe} Fully Connect \num{4096} L1-Memory Banks}
    \label{tab:interconnection}
    \centering
    \input{tables/interconnection}
    \\ \footnotesize\raggedright\textsuperscript{*}\num{4096} \SI{1}{\kibi\byte} \gls{spm} banks are split across each hierarchy with \glspl{pe}, using a banking factor of \num{4}.
    \\ \footnotesize\raggedright\textsuperscript{*}The hierarchy is denoted as $\alpha\text{C}$–$\beta\text{T}$–$\gamma\text{SG}$–$\delta\text{G}$, where $\delta$ is the number of Groups, each with $\gamma$ SubGroups, $\beta$ Tiles per SubGroup, and $\alpha$ \glspl{pe} per Tile.
\end{table*}

\cref{tab:interconnection} presents the comparisons of different hierarchical choices to connect \num{1024} \glspl{pe} to \num{4096} L1 memory banks through multi-stage \gls{fc} crossbar, ranging from a flat (non-hierarchical) to a three-level hierarchy implementation, highlighting the trade-offs between physical feasibility and interconnect performance.

First, the flat design (\emph{1024C}) results in an extremely large \gls{pnr} implementation block, where \num{1024} \glspl{pe} are \gls{fc} to \num{4096} banks through only single crossbar.
Although it offers high throughput and 1-cycle (the lowest) memory access latency, it is physically infeasible due to the unrealistic interconnect complexity.
Introducing the \emph{Tile} hierarchy (\emph{$\alpha\text{C}$–$\beta\text{T}$}) helps mitigate complexity, but the high routing complexity of inter-\emph{Tiles} crossbar still makes the physical routing infeasible.
The \gls{amat} increases because the majority of memory requests are directed to remote \emph{Tiles} and require arbitration at the \emph{Tile} boundaries.
The arbitration rate increases with the number of \glspl{pe} per \emph{Tile}, further reducing interconnect throughput.

Second, by introducing the \emph{Group} level of the hierarchy, we have a two-stage crossbar interconnect (\emph{$\alpha\text{C}$–$\beta\text{T}$-$\delta\text{G}$}), reducing the complexity of critical interconnect nodes by further splitting, and improving physical routing feasibility. 
This choice is a trade-off: it reduces throughput and increases in both \gls{amat} and zero-load latency.
In physical implementation, flattening the \emph{Tiles} into \emph{Group} helps fully leverage the available \gls{beol} resources and simplify interconnect routing across over the \emph{Tiles} without requiring manual port placement.
However, implementing \num{4} large-scale \emph{Groups}, each comprising \num{256} \glspl{pe} and \num{1024} banks, still remains unfeasible due to the significant design effort required for timing and routing optimization, will be described in detail in~\cref{sec:physical_feasibility}.
Additionally, the physically-extended path length limits the achievable design frequency. 
Designing a cluster with more than 4 \emph{Groups} introduces significant floorplan challenges, making it unachievable to arrange the placement of \emph{Groups} to ensure short diagonal access paths while balancing inter-\emph{Group} connections across varying physical distances with same number of pipeline cycles.

Last, we adopt a three-level hierarchical design by dividing each \emph{Group} into \emph{SubGroups} (\emph{$\alpha\text{C}$–$\beta\text{T}$–$\gamma\text{SG}$–$\delta\text{G}$}), while maintaining \num{4} hierarchical blocks ($\gamma=\delta=4$) at each implementation level to ensure balanced diagonal access paths.
To achieve a physically feasible design while maintaining low latency and high throughput, the optimal configuration for the \terapool{} cluster consists of \num{4} \emph{Groups}, each containing \num{4} \emph{SubGroups} of \num{8} \emph{Tiles}, with \num{8} \glspl{pe} per \emph{Tile}, as shown in~\cref{fig:cluster_top}.

\section{TeraPool Cluster Architecture}
\label{sec:cluster}
This section details architecture design elements of \terapool{}, emphasizing the physical design awareness for building the hierarchical implementation from the bottom up.

\subsection{Processing Elements}
\terapool{}'s \glspl{pe} are single-issue, single-stage RISC-V \emph{Snitch} cores based on the \emph{RV32IMA} \gls{isa}~\cite{TeraPool_2024}, where instructions are decoded and executed within the core in a single cycle.
Snitch has a private, fully associative, \gls{scm}-based \num{32}-instruction L0-\gls{IDol}, designed for fast repetition of the computation hot loops.
We build the \emph{Snitch} \gls{cc} with \gls{isa} extensions, including the \emph{Xpulpimg}\footnote{The Xpulpimg extension includes domain-specific instructions~\cite{Mempool_2023}, \eg \gls{mac} and load-post-increment.}  implemented in the \gls{ipu}, and the \emph{zfinx}, \emph{zhinx}, and \emph{smallfloat}~\cite{Fpnew_2024} implemented in the \gls{fpss}.
To address the stringent area constraints typical of many-core compute clusters, we eliminate the need for a costly dedicated \gls{fp} register file and \gls{lsu} by executing \gls{fp} and half-precision operands directly from the integer register file, while supporting parallel operations on two half-types using a \gls{simd} approach.

\begin{figure}[htbp]
  \centering
  \includegraphics[width=\linewidth]{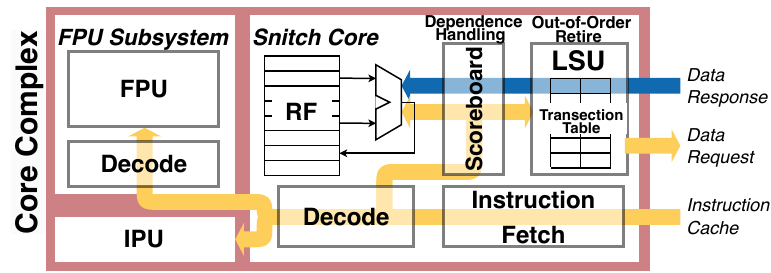}
  \caption{Block diagram of the Snitch highlighting the LSU, and the LSU's transaction table marking outstanding load and stores. }
  \label{fig:cluster_cc}
\end{figure}

On the memory access side, \emph{Snitch} supports non-blocking instruction execution to tolerate multi-cycle L1-memory access latency.
The load and store instructions have their dependencies recorded by the scoreboard, with a transaction table to track their retirement, shown in ~\cref{fig:cluster_cc}.
The outstanding memory requests are tracked in a transaction table, enabling the core to issue a series of loads and stores without blocking while waiting for responses, and eventually proceed to other instructions with no data dependencies.
Given \terapool{}'s \gls{numa} interconnect design, the transaction table retires loads out-of-order, while the scoreboard ensures in-order delivery to the execution units.
By scheduling multiple loads with loop unrolling, we can thus hide the memory access latency within the Cluster's L1 \gls{spm}.
The number of supported outstanding transactions is hardware configurable, and it is chosen at the break-even point where the number of entries avoids excessive area overhead and further performance gains are limited by the number of \gls{isa} registers available.
We found that \num{8} is an adequate number of outstanding transactions for most kernels, for instance, in a blocked \gls{matmul} algorithm~\cite{block_matmul_2014}, a $4 \times 4$ output matrix block—the maximum supported by \num{32} \gls{isa} registers—requires at most \num{8} input transactions.

Snitch's small hardware footprint, latency tolerance, and extensibility make it an ideal candidate for implementing a tunable large-scale shared L1-memory many-core design.

\subsection{Cluster Design}
\begin{figure}[htbp]
  \centering
  \includegraphics[width=\linewidth]{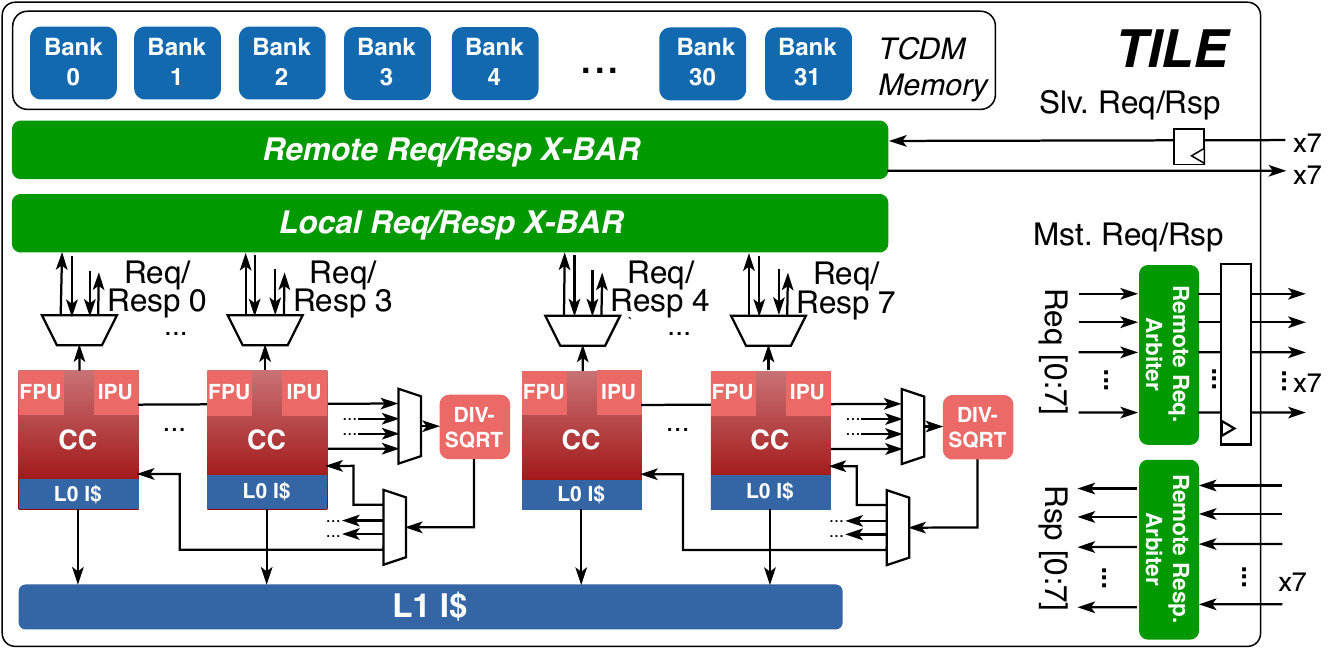}
  \caption{Block diagram of Tile: 8 core-complexes, 2 shared FP-DIVSQRT units; a local crossbar for 1-cycle access to tightly coupled data memory (TCDM), and 7 Remote Req/Resp ports connect to higher-level hierarchies.}
  \label{fig:cluster_tile}
\end{figure}

The \emph{Tile} in~\cref{fig:cluster_tile} is the base hierarchical level, connecting \num{8} \emph{\glspl{cc}} to \num{32} banks of the shared \gls{spm} via a \gls{fc} crossbar, providing single-cycle, zero-load access latency.
The shared \SI{4}{KiB} two-way set-associative L1-\gls{IDol} is shared by all \emph{Snitch} cores within a \emph{Tile} and feeds the L0-\gls{IDol}.
To support \gls{fp} division and square root instructions (essential for operations like matrix inversion but less frequently used than basic \gls{fp} operators), we share a single \gls{fp} \gls{divsqrt} unit among four \emph{\glspl{cc}}, according to a round-robin policy.
Each \emph{Tile} has \num{7} ports to the L1-interconnect: one connects to other \emph{Tiles} within the same \emph{SubGroup} via an \num{8}$\times$\num{8} \gls{fc} crossbar; three connect to \emph{Tiles} in the other \num{3} \emph{SubGroups} within the same \emph{Group} via three \num{8}$\times$\num{8} \gls{fc} crossbars; and the remaining \num{3} connect to \emph{Tiles} in the \num{3} remote \emph{Groups} via three \num{32}$\times$\num{32} \gls{fc} crossbars.
Our interconnect delivers a peak bandwidth of \SI{4}{\kibi\byte\per\cycle} and a bisection bandwidth of \SI{1.875}{\kibi\byte\per\cycle} across hierarchical levels.

\begin{figure}[htbp]
  \centering
  \includegraphics[width=\linewidth]{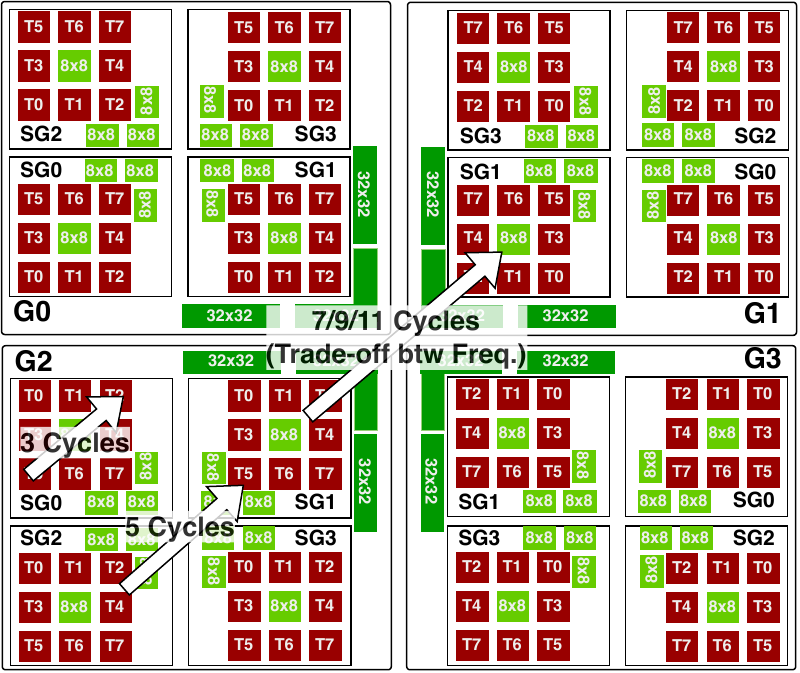}
  \caption{Cluster-level block diagram with layout view of Tiles and interconnect crossbars, highlighting the complexity of each hierarchical crossbar and L1 memory access latency.}
  \label{fig:cluster_top}
\end{figure}

The hierarchical interconnect design provides flexibility in adding spill registers to break long remote access paths.
Spill registers are placed on the \emph{master request} and \emph{response} ports at the issuing and target \emph{Tile} boundaries, as well as after the crossbar on outgoing master ports.
This results in a round-trip access latency of \num{3} cycles within a \emph{SubGroup} and \num{5} cycles for access to other \emph{SubGroups} within the same \emph{Group}. 
The latency for remote-\emph{Group} access is hardware-parameterizable and depends on the target operating frequency. We identify the configurations by subscripts, which indicate the latency of a core access to each hierarchy level. In \terapool{\text{1-3-5-7}}, a register is added on both the request and response paths for each slave port at the \emph{Cluster} level, increasing remote-\emph{Group} access round-trip latency by \num{2} cycles. The \terapool{\text{1-3-5-9}} configuration adds spill registers on the slave ports at the \emph{Group} hierarchy level and the \terapool{\text{1-3-5-11}} configuration adds a register also on the master paths.
The full \emph{Cluster}-level overview with \emph{Tiles} layout is shown in~\cref{fig:cluster_top}.
These configurations represent a trade-off between achievable frequency and L1 worst-case latency, which will be analyzed in~\cref{sec:physical}.

\section{High Bandwidth Memory Link}
\label{sec:system}
We introduce \gls{hbml}, the main memory access interface, which comprises the \gls{dma} engine and system-level global interconnect.
To evaluate \terapool{}'s \gls{hbml} capability, we utilize the open-source, cycle-accurate \emph{DRAMsys5.0} simulator~\cite{DRAMsys_2022} for main memory co-simulation.

\subsection{Hierarchical AXI Interconnect}
\label{sec:axi}
The \gls{hbml} interconnect uses the open and standard \gls{axi} protocol, 
connecting the \terapool{} cluster to main memory and serving as the system-level interface for peripherals, host control, and \glspl{csr}. 
However, providing high-bandwidth connections for each software-managed, physically-addressed L1 \gls{spm} bank poses a significant challenge due to limited physical routing resources, particularly in designs already constrained by \gls{pe}-to-L1 routing complexity.

\begin{figure}[htbp]
  \centering
  \includegraphics[width=\linewidth]{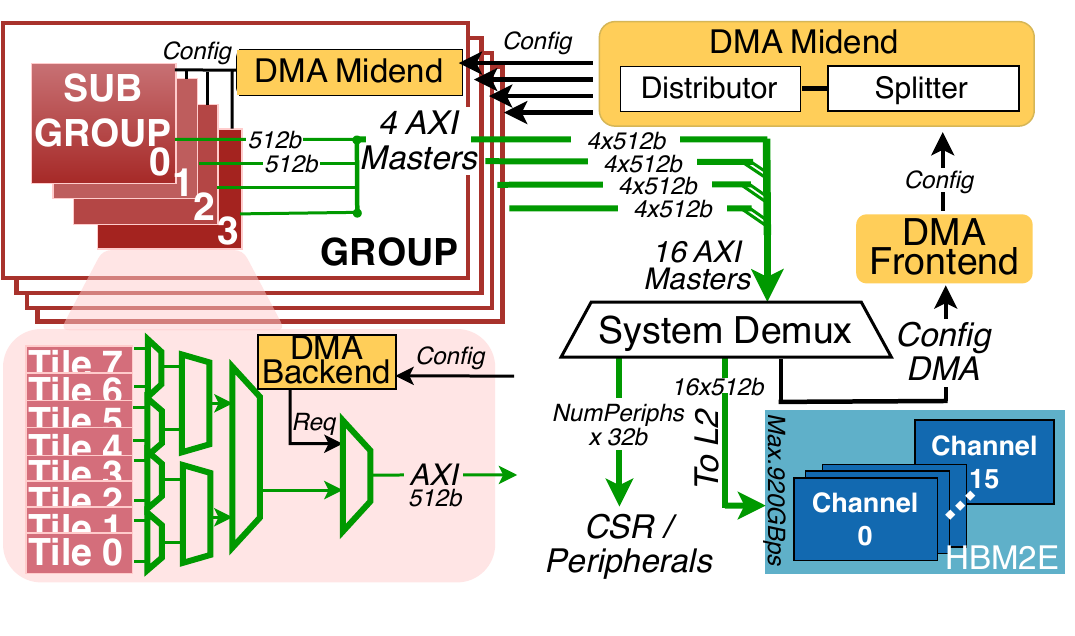}
  \caption{System-level high-bandwidth memory link design, featuring the hierarchical \gls{axi} interconnect and modular \gls{dma} engine.}
  \label{fig:system_hbml}
\end{figure}

To minimize area and routing overhead, we use the tree-like \gls{axi} \gls{hbml} interconnect in~\cref{fig:system_hbml}.
Each \emph{Tile} shares a single \SI{512}{\bit} \gls{axi} master port among its \emph{Snitch} cores.
Core requests are routed through a demultiplexer controlled by the target address, directing them either to the L1 \gls{spm} interconnect or to system-level components.
The \gls{axi} interconnect is also used for L0 \gls{IDol} refills.
Within each \emph{SubGroup}, the \gls{axi} ports from \num{8} \emph{Tiles} are arbitrated in a tree-like structure, connecting to a single \SI{512}{\bit} \gls{axi} master port.
Finally, the \num{16} \gls{axi} masters from each \emph{SubGroup} pass through address-based system demultiplexers, directing them to the \gls{dma} frontend (discussed in~\cref{sec:dma}), L2 main memory or memory controller, and \glspl{csr}/peripheral interfaces.

\subsection{Modular DMA engine}
\label{sec:dma}
Although the \terapool{} cluster distributes its L1 \gls{spm} banks across a large-scale hierarchical design, it maintains a unified physical address space. 
As a result, only one \gls{dma} engine is required to manage software-controlled data transfers between L1 and L2 main memory.
However, utilizing a single traditional \gls{dma} engine presents challenges related to module placement and the need to establish connections to all \num{4096} \gls{spm} banks, resulting in significant routing complexity.
On the other hand, implementing multiple \gls{dma} engines would complicate programming, requiring developers to manage transfer address ranges across different \gls{dma} units.

To simplify the programming interface and reduce routing complexity, we employ a modular iDMA~\cite{iDMA_2024} divided into three components: \emph{frontend}, \emph{midend}, and \emph{backend}.
The \emph{frontend} initiates data transfers by specifying the source and destination address, and transfer size.
Transfer requests are then forwarded to the \emph{midend}, where they are divided into sub-tasks based on the L1 address boundaries of the \emph{SubGroups}.
These sub-tasks are allocated to \gls{dma} \emph{backends}, which manage data transfers between the L2 main memory and the system \gls{axi} interconnect.
Additionally, the \emph{backends} bridge the system-level interconnect and the L1 \gls{spm} crossbar, enabling efficient data delivery to individual banks.

\subsection{Main Memory}
\label{sec:main_memory}
To demonstrate the capability of managing high bandwidth data flows by \terapool{}'s \gls{hbml}, we utilize the open-source \emph{DRAMSys5.0} simulator~\cite{DRAMsys_2022}, a cycle-accurate framework based on \emph{SystemC TLM-\num{2.0}} for \gls{dram} subsystems.
We configure main memory with two \gls{dram} stacks, each with \num{8} \emph{Micron MT54A16G808A00AC-36} \gls{hbm} channels, which support \num{2.8}/\num{3.2}/\SI{3.6}{\giga\bit\per\second}\si{\per pin} \gls{ddr} transmission.

Co-simulation with the \emph{DRAMSys5.0} \gls{hbm} subsystem requires compiling a \emph{C++/SystemC}-based \gls{rtl} cluster model.
The compilation process for a large-scale cluster is extremely time-consuming, particularly during hardware development iterations, where every architectural tuning necessitates recompiling the entire model.
To address this inefficiency, we compile the configured \emph{DRAMSys5.0} model into a \emph{Dynamic Link Library}, compatible with \gls{rtl} simulators such as \emph{QuestaSim} and \emph{VCS}.
This approach significantly accelerates design development, enabling rapid architectural tuning.
We open-source our \emph{DRAMSys5.0} dynamic linkable library simulation environment~\footnote{\url{https://github.com/pulp-platform/dram_rtl_sim}}.

\subsection{Memory Mapping Scheme}
\begin{figure}[htbp]
  \centering
  \includegraphics[width=\linewidth]{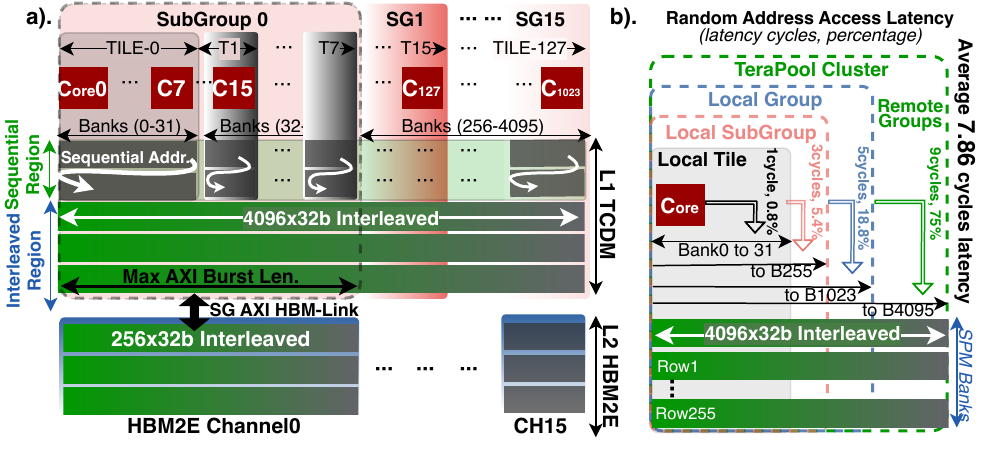}
  \caption{a). Hybrid memory address mapping scheme. b). Access latency when a core randomly accesses L1 memory banks across hierarchy levels.}
  \label{fig:memory_mapping}
\end{figure}

The \terapool{} employs a hybrid addressing scheme, as shown in~\cref{fig:memory_mapping} left.
The \SI{4}{\mebi\byte} L1 \gls{spm} is divided in a \emph{sequential region} and an \emph{interleaved region} (respectively \SI{512}{\kibi\byte} and \SI{3.5}{\mebi\byte} by default, but configurable at design time).
In the \emph{sequential region}, memory requests are kept within a \emph{Tile} to minimize latency and power consumption, allowing programmers to store private data, such as the stack, locally in the \gls{pe}'s \emph{Tile}.
In the \emph{interleaved region}, a word-level interleaved memory mapping is applied across all memory banks.
Data is evenly distributed among all \glspl{pe}, minimizing average access latency and reducing bank conflicts.
The zero-load latency for a core accessing the interleaved address space and the average latency assuming a random access pattern are on the right of \cref{fig:memory_mapping}.
This approach mitigates the performance and power overheads of remote accesses with minimal hardware cost: wire crossings and a multiplexer to scramble the corresponding portion of the request address bits.
Furthermore, interleaving makes access patterns independent of \gls{spm} bank size, which can be flexibly configured at design time.

In each \emph{SubGroup}, the data in the \emph{interleaved region} is distributed across \num{256} banks, with each bank-row storing a \SI{32}{\bit} word.
Thus, when transferring data to or from L2 main memory via the \gls{hbml}, the maximum size of a contiguous address data transfer through a single \gls{axi} burst is \num{256} \SI{32}{\bit} words, handled by the \emph{SubGroup's} \gls{axi} master.
To maintain this maximum burst request length and improve transferring efficiency, we configure \num{1} \gls{dma} \emph{backend} per \emph{SubGroup} to avoid further request splitting.
We align the L2 main memory data interleaving granularity with the \gls{axi} burst length, configuring \num{256} \SI{32}{\bit} words interleaved per \gls{hbm} channel.
This alignment mitigates burst splitting when requested data from a single burst spans multiple \gls{hbm} channels and reduces system interconnect conflicts caused by different \gls{axi} master requests crossing into separate \gls{hbm} channels.

\begin{figure}[htbp]
  \centering
  \includegraphics[width=\linewidth]{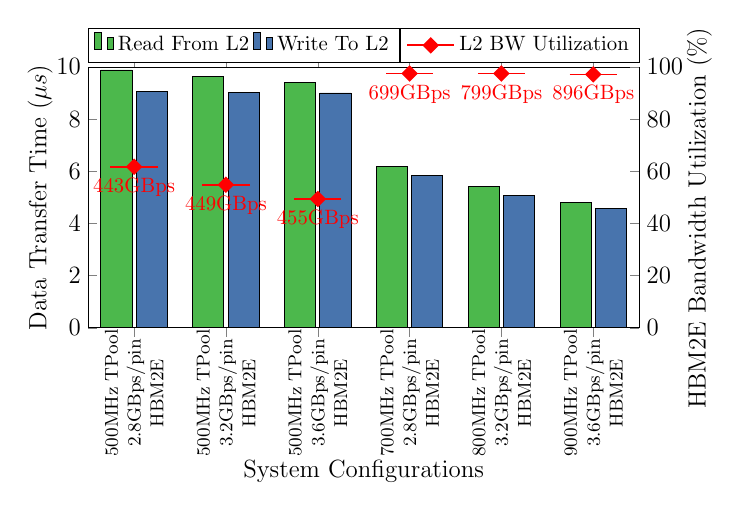}
  \caption{Data transfer performance for TeraPool's L1 read/write of \SI{4}{\mebi\byte} of data, managed by \gls{hbml}, from/to \num{16} \gls{hbm} channels serving as L2 main memory.}
  \label{fig:dram_performance}
\end{figure}

\begin{figure*}[!ht]
  \centering
  \includegraphics[width=\linewidth]{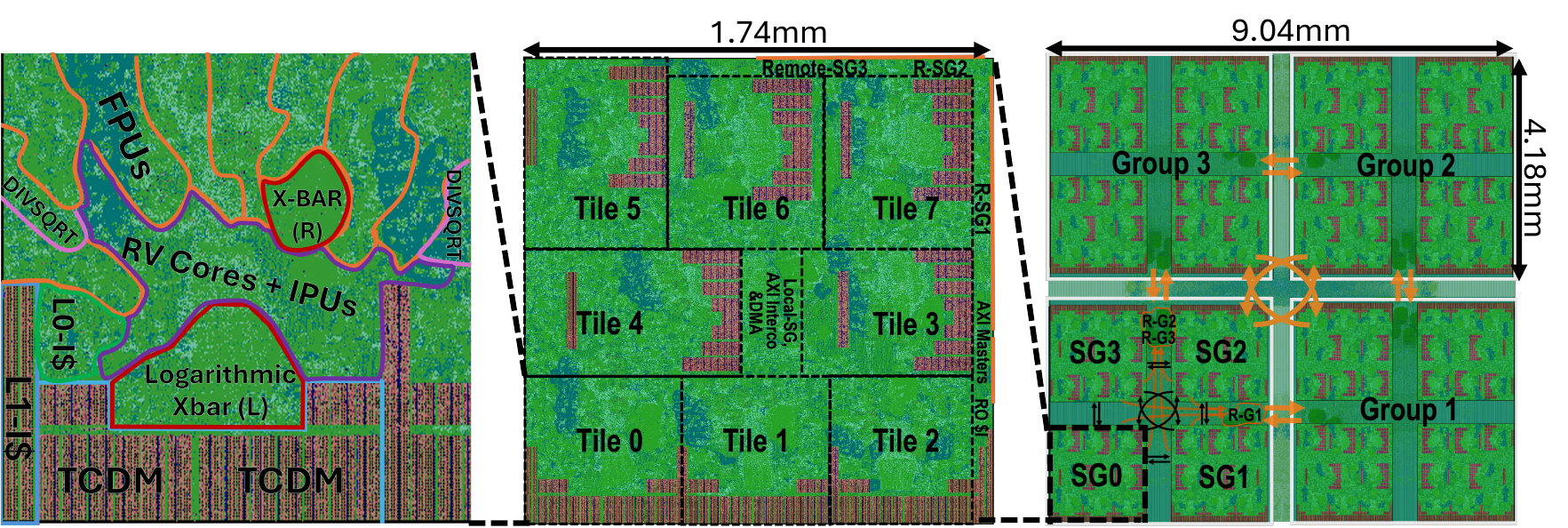}
  \caption{Placed-and-routed layout annotated view of each TeraPool hierarchical instance.}
  \label{fig:physical_layout}
\end{figure*}

To fully utilize the main memory bandwidth (described in~\cref{sec:main_memory}), which supports configurable DDR rates of \num{716.8}-\SI{921.6}{\giga\byte\per\second}, the target implementation frequency of \terapool{} is set to \num{700}-\SI{900}{\mega\hertz} to align with the \gls{hbm} peak bandwidth, leveraging \gls{hbml}'s \num{16}$\times$\SI{512}{\bit} \gls{axi} interfaces.
To validate this analysis, we explore \terapool{} at both \SI{500}{\mega\hertz} and \SI{900}{\mega\hertz}, performing intensive data transfers (input\&output) across the full \SI{4}{\mebi\byte} cluster memory under various \gls{hbm} DDR rate configurations.
The results are shown in~\cref{fig:dram_performance}.
At the lower frequency target (\SI{500}{\mega\hertz}), bandwidth is constrained by \terapool{}'s operating frequency, utilizing \SI{61.8}{\percent}-\SI{49.4}{\percent} of peak \gls{hbm} bandwidth. In contrast, at \num{700}-\SI{900}{\mega\hertz}, all three corresponding \gls{hbm} DDR configurations achieve near-peak utilization (\SI{97}{\percent}), reaching up to \SI{896}{\giga\byte\per\second} at \SI{900}{\mega\hertz}.
The slight bandwidth loss is attributed to \gls{dma} \emph{frontend} configuration cycles and unavoidable \gls{dram} refresh periods.
This analysis further motivates \terapool{} design target to near-\si{\giga\hertz} frequency, and demonstrates \terapool{}'s \gls{hbml} capability to efficiently manage high-bandwidth, long-latency data transfers, even with industry-standard \gls{hbm} exhibiting hundred-cycle latencies as main memory.

\section{Physical Implementation}
\label{sec:physical}
In this section, we present the details of \terapool{}'s physical design.
The \terapool{} cluster is implemented within a die area of \SI{81.8}{\milli\meter\squared} using \emph{GlobalFoundries'} \SI{12}{\nano\meter} LP+ FinFET process with a \num{13}-metal layer stack.
We use \emph{Synopsys' Fusion Compiler 2022.03} to synthesize, place, and route.
Our multi-level \gls{fc} interconnect design is well-suited to the bottom-up implementation flow: it defines clear boundaries for each hierarchical block and allows the addition of spill registers for timing control.
We determine the power consumption under typical operating conditions (TT/\SI{0.80}{\volt}/\SI{25}{\celsius}) using Synopsys' PrimeTime 2022.03 and referring to switching activities obtained from a post-layout gate-level simulation.

\subsection{Design Feasibility and Floorplan}
\label{sec:physical_feasibility}
\begin{figure}[htbp]
  \centering
  \includegraphics[width=\linewidth]{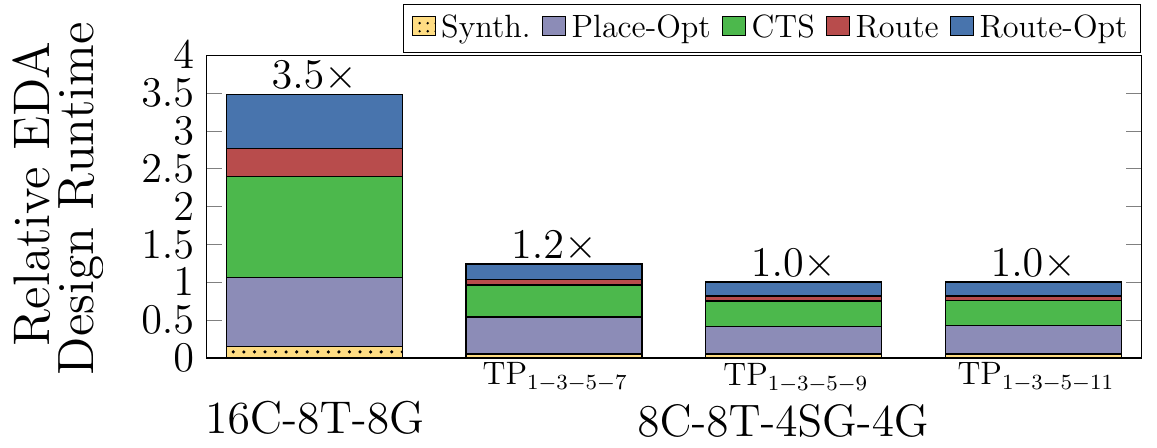}
  \caption{Breakdown of the relative implementation time for TeraPool’s Group in different design configurations.}
  \label{fig:eda_runtime}
\end{figure}

To validate our analysis in~\cref{sec:interco_analysis} of the substantial \gls{eda} efforts required for large-scale base-block implementation, which result in unmanageable runtimes and degraded optimization performance, we implement the \emph{\num{16}C-\num{8}T-\num{8}G} hierarchical configuration (refer to~\cref{tab:interconnection}).
\cref{fig:eda_runtime} presents the relative \gls{eda} tools' runtime breakdown for the implementation of a \terapool{} Group.
For the non-implementable \emph{\num{16}C-\num{8}T-\num{8}G} hierarchical configuration, the total relative \gls{eda} tools' runtime is nearly $3.5\times$ that of \terapool{\text{1-3-5-9}}, with timing optimization accounting for more than \SI{80}{\percent} of the effort.
Specifically, the routing stage is $5.5\times$ slower than the other configurations.
Despite these high design optimization efforts, the eight $16\times16$ interconnects in the Group design introduce significant routing congestion, leading to numerous metal shorts.
Furthermore, routing detours considerably increase the length of timing paths, making it impossible to close the design with a \SI{500}{\mega\hertz} target (TT/\SI{0.80}{\volt}/\SI{25}{\celsius}).
These findings fully support our insights in~\cref{sec:interco_analysis} regarding \terapool{}'s physically-aware hierarchical interconnect analysis.

\cref{fig:physical_layout} shows a snapshot of the physical layout with annotated module placement.
The Subgroup is implemented in \SI{3.03}{\milli\meter\squared}, resulting in \SI{0.047}{\milli\meter\squared\per\core} and a satisfactory \SI{58}{\percent} area utilization.
Each Subgroup comprises \num{8} Tiles, with the \gls{spm} macros of each Tile grouped and placed in a U-shape to enclose the local crossbar, thereby minimizing the overall distance and avoiding excessive macro stacking.
The center and edges of the Subgroup are left free to place interconnect cells, and the macros U-shape is rotated on the bottom side to better utilize the design area.
The interface ports for remote-Subgroups and remote-Groups are placed on the northeast boundaries.
Subgroup and Group blocks are arranged in a point-symmetric grid, with channels in between to \gls{pnr} the interconnects.
To further improve area utilization, we place interface ports behind the Subgroup blocks and shrink the channel width until \gls{beol} resources become limited.
At the Cluster top level, \SI{0.68}{\milli\meter}-wide channels are allocated for connecting Groups' crossbars across the center, and for linking AXI-to-\gls{hbm} channels to boundaries.
These routing channels increase the overall area per core to \SI{0.079}{\milli\meter\squared\per\core}.

\subsection{Performance and Logic Area}
We configure and implement hardware-parameterizable latency for remote Group access with \num{7}-/\num{9}-/\num{11}-cycle, achieving operating frequencies of \num{730}-/ \num{850}-/\SI{910}{\mega\hertz}, respectively (TT/\SI{0.80}{\volt}/\SI{25}{\celsius}).
These results clearly illustrate the trade-off between latency and operating frequency.
When the configured latency exceeds \num{11} cycles, the design speed is no longer constrained by the remote Group access path, but by the Snitch core:
It starts at a register after the instruction cache, passes through Snitch and a request interconnect, and arrives at the clock gating of an \gls{spm} bank.

\begin{figure}[htbp]
  \centering
  \includegraphics[width=\linewidth]{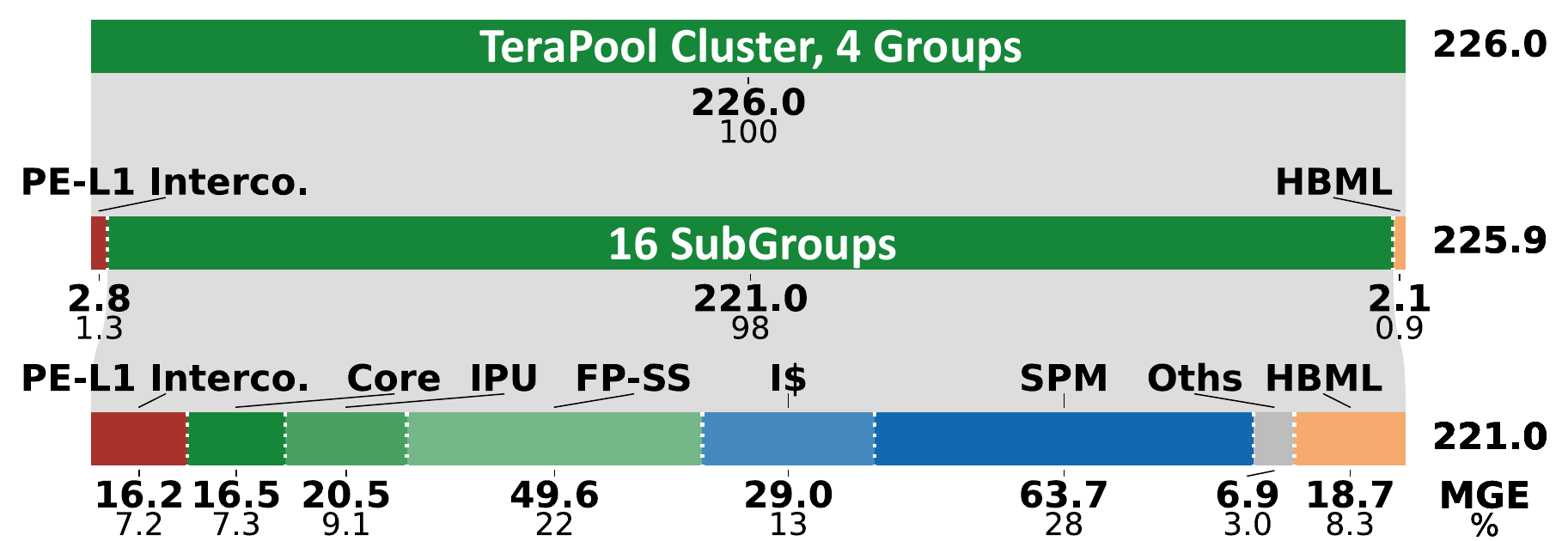}
  \caption{Hierarchical area breakdown with annotations showing the \emph{percentage} of the immediate parent component.}
  \label{fig:area_breakdown}
\end{figure}

The \gls{ge}, defined as the area of a \num{2}-input \emph{NAND} gate, represents a technology-independent metric for logic area.
As shown in~\cref{fig:area_breakdown}, our hierarchical interconnect design accounts for only \SI{8.5}{\percent} and the \gls{hbml} occupies a negligible \SI{9.2}{\percent} of the cluster area.
The most significant area component is the \gls{spm} banks, followed by the \emph{Snitch \glspl{cc}} and the instruction cache.
The \emph{\glspl{cc}} can be further divided into the cores themselves (\SI{7.3}{\percent}), \glspl{ipu} (\SI{9.1}{\percent}), and the highly area-efficient \glspl{fpss} (\SI{22}{\percent}).
This breakdown highlights that the majority of the design’s logic area is dedicated to memory and compute resources, with minimal area cost attributed to both \glspl{pe}-to-L1-memory interconnects and \gls{hbml}.

\begin{figure*}[!ht]
  \centering
  \includegraphics[width=\linewidth]{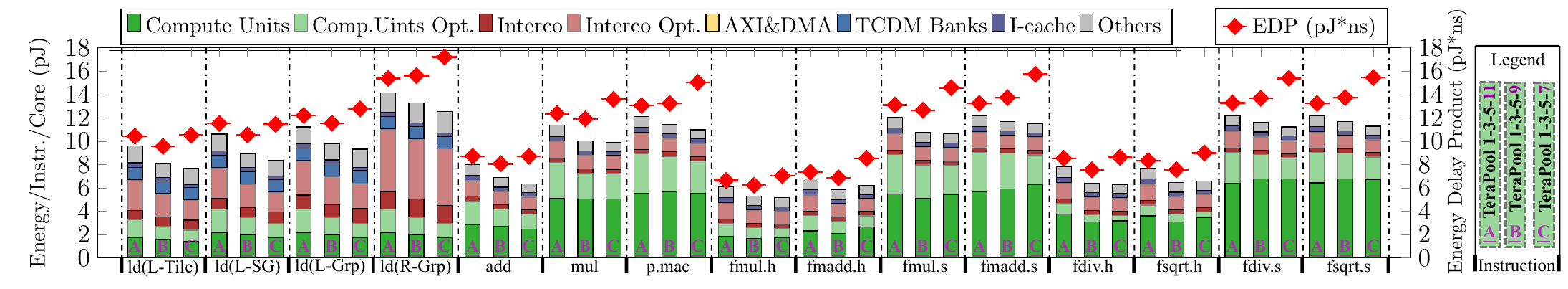}
  \caption{Instruction energy breakdown under typical \SI{0.80}{\volt}/\SI{25}{\celsius} for varying TeraPool configurations (7/9/11 remote Group access latency cycles). Memory access instructions (ld) target local Tile, SubGroup, Group, and remote Group; floating-point instructions presented in single (.s) and half-precision (.h); division/square root energy quantified per DIVSQRT unit.}
  \label{fig:instruction_energy}
\end{figure*}

\subsection{Energy Analysis}
\cref{fig:instruction_energy} shows the energy breakdown for instruction execution (\si{\pico\joule\per instruction\per core}) in \terapool{} configured with remote \emph{Group} access latencies of \num{7}-, \num{9}-, and \num{11} cycles, measured at corresponding operating frequencies of \num{730}-, \num{850}-, and \SI{910}{\mega\hertz} (TT/\SI{0.80}{\volt}/\SI{25}{\celsius}).
To demonstrate the energy consumption across different memory access distances, we present the energy breakdown for a \SI{32}{\bit} load instruction (\emph{ld}) accessing memory banks within its \emph{local-Tile}, a different \emph{Tile} in the same \emph{SubGroup}, a different \emph{SubGroup} in the same \emph{Group}, and a remote \emph{Group}.
With increasing memory access distance, the average energy consumption rises by \SI{10}{\percent}, \SI{20}{\percent}, and \SI{58}{\percent} compared to the \emph{local-Tile}, respectively.
The interconnect (\SI{2.5}-\SI{6.8}{\pico\joule}) and L1 \gls{spm} banks (\SI{1.06}{\pico\joule}) dominate the energy consumption, contributing up to \SI{51}{\percent} of the total.
For arithmetic instructions, \terapool{} consumes \num{6.4}–\SI{13.5}{\pico\joule} for integer operations, \num{5.2}–\SI{7.9}{\pico\joule} for half-precision (\SI{16}{\bit}) \gls{fp}, and \num{11.3}–\SI{12.2}{\pico\joule} for single-precision (\SI{32}{\bit}) \gls{fp}.
The computing units (\gls{ipu}, \gls{fpss}, and \gls{divsqrt}) dominate energy consumption, accounting for an average of \SI{69.9}{\percent} in integer, \SI{60}{\percent} in half-precision \gls{fp}, and \SI{72.3}{\percent} in single-precision \gls{fp} computations.
When the interconnect crossbars are not accessed, for instance, during the execution of a single-precision FMA (\gls{fp} Multiply-Add, \texttt{fmadd.s}) operation, which consumes \SI{12.19}{\pico\joule}, only \SI{14.5}{\percent} of the total power, primarily due to internal cell power from clock propagation and leakage.
Thanks to memory clock gating, the \gls{spm} banks consume less than \SI{0.1}{\pico\joule}, achieving a \SI{98}{\percent} reduction in energy consumption and becoming negligible when not accessed.

The light-colored segments in the plot, adjacent to the \emph{Compute Units} and \emph{Interconnect} segments, represent the energy consumption of optimization cells introduced by physical design to achieve the target frequency.
These segments show a slight increase as the frequency rises, driven by the use of low-threshold optimization cells, designed to minimize delay at the expense of higher power consumption, to meet the near-gigahertz operating frequency target.
For instance, as the operating frequency increases from \SI{730}{\mega\hertz} to \SI{910}{\mega\hertz} under typical conditions (TT/\SI{0.80}{\volt}/\SI{25}{\celsius}), the energy consumption rises by only \SI{1.6}{\pico\joule} for remote \emph{Group} memory access and \SI{1.4}{\pico\joule} for single-precision \gls{fp} multiplication (\emph{fmul}), reflecting an average increase of \SI{16}{\percent}.

The \gls{edp} is a key metric that normalizes energy efficiency and delay~\cite{manycore_edp_2018}, providing insight into the trade-offs between energy consumption and performance.
To identify the optimal energy-performance configuration for \terapool{}, we analyze the \gls{edp} in $\si{\pico\joule} \times \si{\nano\second}$ for each instruction, highlighted with red markers in~\cref{fig:instruction_energy}.
The results clearly show that, for most operations, the \terapool{} configuration with a \num{9}-cycle remote \emph{Group} latency operating at \SI{850}{\mega\hertz} achieves the optimal balance between energy efficiency and performance.
In summary, \terapool{} achieves an energy-efficient design, consuming only \SI{5}{\sim}\SI{15}{\pico\joule\per operation\per core}, even for high-complexity \SI{32}{\bit} \gls{fp} operations, the majority of the energy is spent toward computing components.

\section{Performance Analysis}
\label{sec:software}
One advantage of a shared-memory cluster is its programming simplicity compared to managing data transfers across multiple clusters.
\terapool{} supports a streamlined \emph{fork-join} programming model with C-runtime functions, which enable efficient transitions between sequential and parallel code sections for \gls{spmd} execution~\cite{Programming_Model_2012}.
After booting all the \glspl{pe} execute concurrently the C-binary, the programmer can access \glspl{pe}' unique ID and statically assign to each core independent tasks in a parallel computation (\emph{fork}).
Once \glspl{pe} complete, they synchronize (\emph{join}) via atomic \emph{fetch\&add} to shared memory locations.

To evaluate interconnect performance and overall cluster utilization, the benchmarks cover data-parallel execution with both sequential and non-sequential memory access patterns.
For sequential memory access execution, \gls{axpy} and \gls{dotp} are used as \emph{local-access} kernels, where \glspl{pe} primarily fetch data from nearby regions of the shared memory with low latency, and inputs are not fetched by more than one core, making them ideal for evaluating the lower-hierarchy interconnect.
The \gls{gemm} serves as a \emph{global-access} kernel,  where data structures are distributed across all shared \gls{spm} banks.
A tiled implementation is used to fully utilize the register file and maximize computational intensity.

For non-sequential memory accessing, we adopt a radix-4 decimation in frequency Cooley-Turkey \gls{fft} to ease the memory accesses in local banks.
In the $k^{th}$ stage of an N-point \gls{fft}, each core computes \num{4} butterflies, taking \num{4} inputs at a distance of $N/(4 \times 4k)$.
\glspl{pe} working on different \glspl{fft} are independently synchronized.
We also include \gls{spmmadd} with \gls{csr} format, which performs element-wise addition of two graphs and represents a key \emph{GraphBLAS} operation~\cite{blas_2016}, to evaluate the interconnect under irregular distributed memory accesses.

\begin{figure}[!htbp]
  \centering
  \includegraphics[width=\linewidth]{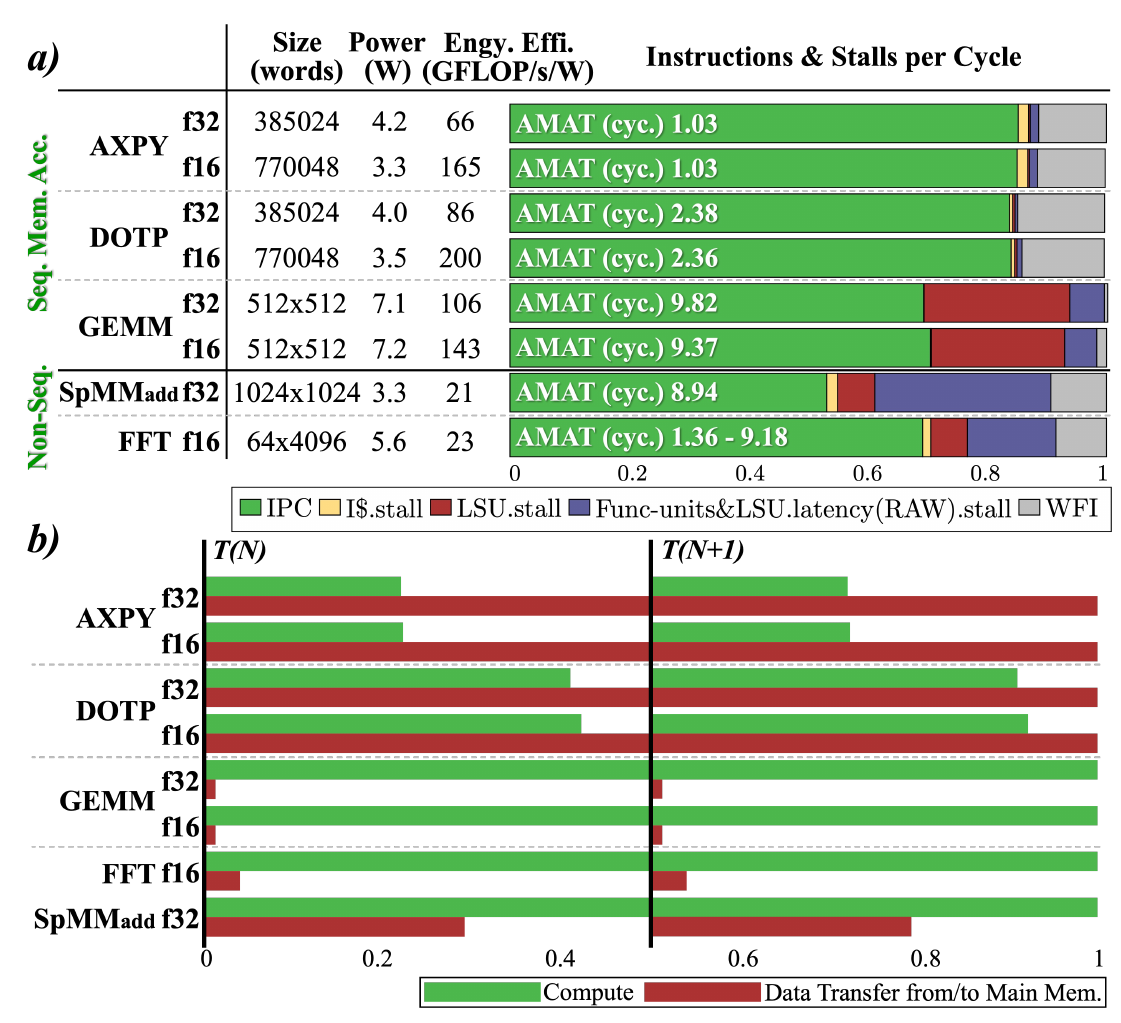}
  \caption{a) Instruction and stall fractions over total cycles for kernel execution in \terapool{}; b) Timing breakdown of double-buffered kernels with data transfer to/from \gls{hbm}.}
  \label{fig:IPC}
\end{figure}

In~\cref{fig:IPC} a), we present kernel performance on the most energy-efficient configuration with a \num{9}-cycle remote Group latency, operating at \SI{850}{\mega\hertz} (TT/\SI{0.80}{\volt}/\SI{25}{\celsius}).
We break down the \gls{ipc} and architectural stall fractions as a percentage of total cycles.
Since all memory accesses are local, both \gls{axpy} and \gls{dotp} achieve low \gls{amat} and high \gls{ipc} values of \num{0.85} and \num{0.83}, respectively, benefiting from the high bandwidth of the local-\emph{Tile} interconnect.
The performance loss is attributed only to the synchronization barrier (\gls{wfi}) at the end of the computation.
\gls{dotp} incurs more \gls{amat} and synchronization overhead due to the reduction.
For \gls{gemm}, a global-access-dominated kernel where cores fetch data across the entire cluster, the measured \gls{amat} aligns closely with the random-access analytical model presented in~\Cref{sec:interconnection}, validating model accuracy for design guidance.
The \gls{ipc} remains high at \num{0.7}, while a larger fraction of \gls{lsu} stalls is due to global accesses that require \glspl{pe} to fetch data from all banks through shared interconnect ports.
A smaller fraction of \gls{raw} stalls arises from increased latency during interconnect conflicts.
The negligible synchronization overhead, thanks to our large shared memory, enables the cluster to handle larger problem tiling efficiently.

For non-sequential memory accessing, we run \num{64} independent \num{4096}-point \glspl{fft} in parallel, computing each stage between barriers to minimize overhead.
It achieves \num{0.7} \gls{ipc}, performing at \SI{23}{\giga\flop\per\second\per\watt} due to the required data-control instructions such as packaging and shuffling alongside computation.
The \gls{amat} ranges from \num{1.36} to \num{9.18} cycles across different stages.
Data access through interconnects within and between \emph{SubGroups} incurs only \SI{6.15}{\percent} \gls{lsu} stalls.
Although \glspl{pe} are not optimized with dedicated extensions for sparse workloads, \gls{spmmadd} still achieves a high \gls{ipc} of \num{0.53}. 
Frequent conditional branches stress the non-specialized \glspl{pe}, where the limited number of \gls{isa}-defined registers hinders loop unrolling and leads to \gls{raw} stalls.
However, when analyzing memory accesses, we note that non-sequential narrow accesses incur only \SI{6}{\percent} interconnect contention, demonstrating our interconnect supports multiple irregular parallel accesses in a non-blocking fashion.

To assess the high bandwidth of TeraPool's \gls{hbml}, we benchmark kernels accessing data through industry-standard \gls{hbm}, employing double-buffering to hide transfer latency.
We use two L1 buffer spaces: one for executing the current kernel (\emph{T(N)}) and another for transferring data for the next round (\emph{T(N+1)}).
As shown in~\cref{fig:IPC} b), even for heavily memory-bound kernels, \gls{dotp} achieves a compute phase of \SI{82}{\percent}, while \gls{axpy} achieves \SI{44}{\percent}, as fast computation fails to hide the latency of result transfers and new data loads.
For compute-bound kernels, \terapool{} achieves high performance: its large shared-L1 completely hides the \gls{hbm} latency.

These analyses highlight the capability of our \gls{pe}-to-L1 interconnect and \gls{hbml} in delivering a significant fraction of the peak workload, even for kernels with highly global \gls{pe}-to-L1 traffic patterns, non-negligible contention, and long-latency data transfers with \gls{hbm} main memory.

\section{Related Work}
\label{sec:soa}
In this section, we compare our work with cluster-based, programmable, multi-core and many-core architectures that employ various scaling-up/-out topologies to achieve high core-counts.
We privilege algorithmic flexibility, and focus on clusters of general-purpose \glspl{pe}, in contrast to dense arrays of hyper-specialized, single-domain \glspl{pe} (e.g., \glspl{npu}).
\cref{tab:SoA} provides an overview of these architectures.
We can divide the state-of-the-art into two main categories: multi-core designs with a cluster's core count $<100$ and many-core designs with $>100$.

In the multi-core designs category, we find both scaled-up/-out topologies:
\viewpoint{1} crossbar-based clusters scaling-out via a 2-D mesh \gls{noc} with router links such as \emph{Kalray MPPA3-80}~\cite{Kalery_2021}, and
\viewpoint{2} scaled-up clusters with crossbar-based interconnects, such as \emph{Ramon RC64}~\cite{Ramon_2016,Ramon_2021}. In the latter, all cores connect to a shared \SI{4}{\mega\byte} \gls{spm} employing multistage logarithmic crossbars. The core count is limited to reduce routing complexity.
Our work clearly outperforms these designs by virtue of a much larger core count, achieving \num{32}-\num{128}\si{\times} performance.

In the many-core design category, among the scaled-out architectures, \emph{TensTorrent's Wormhole}~\cite{Tenstorrent_2021} integrates \num{400} \glspl{pe} in a $10 \times 8$ \emph{Tensix Matrix} interconnected via a 2-D mesh \gls{noc}.
In each \emph{Tensix} only \num{5} \glspl{pe} share \SI{1.5}{\mega\byte} L1 cache through a crossbar with \num{4} pipeline stages.
With the same round-trip pipeline, \terapool{}'s \emph{Group} features \num{256} \glspl{pe} interconnected to the \emph{Group}-local portion of a \SI{1}{\mebi\byte} shared-memory, achieving a $76.8\times$ \glspl{pe}-to-local-shared-memory ratio, thanks to our single-stage \gls{pe} design with a fully combinational crossbar, incorporating only spill registers at hierarchical boundaries to break long remote access paths.
In \emph{Esperanto ET-SoC-1}~\cite{ET_soc_2022}, a crossbar connects \num{32} \emph{Minion} cores to a \SI{4}{\mega\byte} L1 cache, scaling out to \num{1088} cores through a 2-D mesh \gls{noc} while utilizes \gls{rvv} \glspl{pe} with a large vector register file to hide transaction latency.
\terapool{} achieves a similar core count by only scaling up, with latency hidden by a negligible-sized transaction table instead of relying on a power-costly large register file in our lightweight scalar core design.
Our design achieves $16\times$ higher L1 bandwidth and $32\times$ higher L2 bandwidth.
\emph{HammerBlade}~\cite{Hammerblade_2024} scales out from a single-core \emph{Tile} with \SI{4}{\kilo\byte} \gls{spm}, using a mixed 2D-mesh topology: \emph{Tiles} are connected via Half-Ruche channels horizontally, direct mesh links vertically, and wormhole-skipped links between \emph{Cache Tiles} and HBM2 for cache refilling.
The shared \gls{spm} region (\emph{Tile Group}) is defined at design time within a 128-core \emph{Cell} (up to \SI{0.5}{\mebi\byte}), which is \num{8}$\times$ smaller than our shared-memory scale.
Despite this sophisticated topology, accessing the shared memory can take up to \num{52} cycles and requires each \emph{Tile} to support up to \num{63} outstanding requests.
In contrast, TeraPool achieves a \num{46}$\times$ advantage in the ratio of shared-memory size to maximum access latency.
As a result, \emph{HammerBlade} demands data-placement-aware programming to mitigate latency penalties, increasing programming complexity.

The \emph{Occamy}~\cite{Occamy_2025} scales out from a cluster where \num{8} RISC-V \glspl{pe} share a low-latency (\SI{1}{cycle}) \SI{128}{\kibi\byte} L1 memory through a crossbar-based interconnect.
To enable parallel processing on large datasets, \num{24} scaled-out clusters are interconnected, with data managed by \glspl{dma} within each cluster and inter-cluster interconnects occupying \SI{14.8}{\percent} of the die area.  
In contrast, \terapool{} eliminates inter-cluster interconnects, featuring a $32\times$ larger L1 size managed by a single \gls{dma}, further simplifying programming.
Finally, in modern \glspl{gpu}' \glspl{sm}, \glspl{pe} share memory through a crossbar-based interconnect.
For instance, \emph{NVIDIA’s H100}~\cite{nvidia_h100_2023,nvidia_h100_uncovering_2024} is organized into \glspl{sm} with \num{128} \si{FP32} units sharing \SI{256}{\kilo\byte} memory, and scales-out via a data-driven \gls{noc} mesh to a total of \num{18432} \glspl{pe}.
In \terapool{}, the shared memory is $16\times$ larger, with $4\times$ L1 bandwidth per \gls{pe}, while remaining physically feasible.
Moreover, to support \gls{simt} execution, each \gls{sm} incorporates multiple schedulers, dispatch units, and large register files, significantly increasing hardware complexity.  
In contrast, \terapool{}'s low-latency shared memory with outstanding transaction support avoids reliance on thread switching to hide latency, reducing hardware costs.
Additionally, its \gls{spmd} execution employs a unified multi-megabyte L1 address space, eliminating the need for shared data replication across separate memory regions.

To our knowledge, the largest scaled-up shared-memory cluster in the literature is MemPool~\cite{Mempool_2023}.
It follows a design philosophy similar to our work, featuring \num{256} cores sharing a \SI{1}{\mebi\byte} \gls{spm}.  
With our proposed hierarchical interconnect and physical-design-aware methodology, we scale up the cluster by $4\times$, incorporating more powerful \glspl{pe} with multi-precision \gls{fp} support.  
Physical feasibility is maintained, enabling all \glspl{pe} to access any of the \num{4096} shared memory banks with only a \num{2}–\num{6} (depending on target frequency) cycle latency increase at most, which is effectively hidden by the \gls{pe}'s non-blocking load/store unit and proper instruction scheduling.

It is possible to quantitatively and fairly compare \terapool{} with Occamy and MemPool since they are all open-source architectures, representative of small (tens of \glspl{pe}), large (hundreds of \glspl{pe}), ultra-large (thousand \glspl{pe}) clusters.
\cref{tab:transfer_vs_ipc} compares the main memory accesses required to sustain the workload per unit computation (Byte/FLOP) and utilization (\gls{ipc}) for two key kernels, namely \gls{axpy} and \gls{gemm}.
Clearly, we want to minimize the former and maximize the latter.
For a kernel with low data reuse like \gls{axpy}, the Byte/FLOP remains unchanged across designs.
\terapool{}, the largest cluster design, achieves the same high utilization (\num{0.85}~\gls{ipc}), as data can be stored close to \glspl{pe}, to leverage the high-bandwidth of low-latency local interconnects.
For \gls{gemm}, we see a more complex trade-off: \gls{ipc} decreases for \terapool{}, due to increased latency in the remote \emph{Group} interconnects, and the conflicts at the hierarchy boundaries when cores within the same \emph{Tile} concurrently access the same \emph{Group}.
Nevertheless, the \gls{ipc} remains high at \num{0.7}.
On the other hand, the large capacity of \terapool{}'s memory enables large-scale tiling and more data reuse.
The \SI{19}{\percent} drop in \gls{ipc} is more than compensated by the \SI{44}{\percent} and \SI{85}{\percent} reduction of system-level bandwidth-per-core required to sustain the workload, especially in view of the well-known "\emph{Memory Wall}" problem.

\begin{table*}
    \caption{Comparison of \gls{soa} Cluster-based Many-core Designs}
    \label{tab:SoA}
    \centering
    \resizebox{\textwidth}{!}{%
        \input{tables/SoA}
    }
    \\ \footnotesize\raggedright\textsuperscript{a} In 32-bit operations, one MAC is counted as two operations.
       \footnotesize\raggedright\textsuperscript{b} A \num{4}-cycle interconnect pipeline with unknown core latency.
       \footnotesize\raggedright\textsuperscript{c} Benchmarking referenced from \url{https://chipsandcheese.com/p/nvidias-h100-funny-l2-and-tons-of-bandwidth}, last accessed on Feb. 28, 2025.
       \footnotesize\raggedright\textsuperscript{d} We assume one \emph{Tile group} per HB \emph{Cell} as a shared-memory cluster, where the \emph{Group} SPM address space is shared across all \emph{Tiles}.
       \vspace{-1em}
\end{table*}

\begin{table}[!ht]
    \caption{Data Transfer Cost vs. Compute IPC}
    \label{tab:transfer_vs_ipc}
    \centering
    \resizebox{\columnwidth}{!}{
        \input{tables/transfer_vs_IPC}
    }
    \\ \footnotesize\raggedright\textsuperscript{*} Same configuration of \gls{pe}, transaction table, and instruction cache.
    \\ \footnotesize\raggedright\textsuperscript{**} \SI{8}{\mebi\byte} problem matrices in main memory; tiling based output matrix.
\end{table}

\section{Cluster Scale-Up Trade-Offs and Future Work}
\label{sec:future_work}
While we demonstrate the hardware and software benefits of larger clusters, scaling up introduces two main trade-offs in \gls{pe}-to-L1-memory interconnects.
The first is the floorplan area.
Although our interconnect ensures physical feasibility, it requires routing channels (\SI{40}{\percent} die area) between implementation blocks for crossbar placement and wiring.  
In contrast, smaller clusters that follow the same design principle, such as MemPool~\cite{Mempool_2023}, incur lower area overhead due to fewer hierarchical levels.
The second trade-off concerns inter-hierarchy bisection bandwidth scaling.
A higher peak-to-bisection bandwidth ratio ($\mathit{peakBW}/\mathit{bisecBW}$) indicates more contention when memory accesses are not localized.  
Compared to MemPool (\SI{1}{\kibi\byte\per\cycle} peak, \SI{0.75}{\kibi\byte\per\cycle} bisection), TeraPool scales by \SI{4}{\times} with a linearly increased peak bandwidth of \SI{4}{\kibi\byte\per\cycle}.
However, the bisection bandwidth reaches only \SI{1.875}{\kibi\byte\per\cycle}, constrained by wiring resources, increasing the peak-to-bisection ratio by \SI{60}{\percent}.

To address these trade-offs, future work will focus on exploring more area-efficient interconnect designs.
A promising direction is using 2D-mesh \glspl{noc} with over-macro routing~\cite{Mempool_2023} for \gls{pe}-to-L1-memory communication.  
The mesh regular wiring topology allows exploiting controlled routing on higher metal layers and reduces channel overhead, and the configurable router count allows fully utilizing wiring resources to improve bisection bandwidth.
However, as the end-to-end latency increases with hop count, \glspl{noc} are less suitable for latency-sensitive core-to-L1-memory access.
Combining 2D-mesh networks with crossbars may provide a balanced solution that supports scalability and low latency.

Another direction of evolution is application-oriented architectural customization.  
While this work presents TeraPool with general-purpose \glspl{pe}, future deployments may benefit from dedicated \gls{isa} extensions or co-processors tailored to specialized workloads and data formats (e.g., sparsity and quantized data formats).
Moreover, the local-Tile folded sequential \gls{spm} address regions are well-suited for integrating specialized engines, such as tensor arrays or \glspl{npu}, to accelerate AI-centric computation.

\section{Conclusion}
\label{sec:conclusion}
\terapool{} is a physically feasible many-core cluster design featuring \num{1024} individually programmable, floating-point-capable RISC-V cores sharing \SI{4}{\mebi\byte} of L1 \gls{spm} through \gls{numa} hierarchical crossbar interconnects delivering up to \SI{4096}{\byte\per\cycle} bandwidth.
The \SI{932}{\giga\byte\per\second} High-Bandwidth Memory Link is designed to efficiently manage data transfers in/out of the large shared L1.
It improves by $\times$\num{4} the \gls{pe} count with respect to the largest state-of-the-art cluster.

Through a comprehensive analysis of the core-L1 interconnect, we successfully placed and routed the cluster using \emph{GlobalFoundries}' 12LPPLUS FinFET technology.
TeraPool delivers \num{0.27}-\SI{0.74}{\tera\flop\per\second} single-precision, and up to \SI{1}{\tera\flop\per\second} half-precision performance, achieving \SI{23}{\giga\flop\per\second\per\watt} to \SI{200}{\giga\flop\per\second\per\watt} energy efficiency on real-world kernels.
This work demonstrates that the shared-L1 memory paradigm can scale to a thousand cores while maintaining physical feasibility, providing substantial parallel computing power.
The unified-address shared-L1 memory accommodates multi-megabyte data chunks while eliminating the cost of inter-cluster global interconnects, minimizing data transfer, splitting-merging, and synchronization overhead.

\ifCLASSOPTIONcompsoc{}
  \section*{Acknowledgments}
\else
  \section*{Acknowledgment}
\fi

This work has received funding from the Swiss State Secretariat for Education, Research, and Innovation (SERI) under the SwissChips initiative.

\Urlmuskip=0mu plus 1mu\relax
\def\UrlBreaks{\do\/\do-}
\bibliographystyle{IEEEtran}
\bibliography{bib/terapool}

\input{files/biography}

\end{document}

%% file: files/packages.tex
\ifCLASSOPTIONcompsoc%
  \usepackage[nocompress]{cite}
\else
  \usepackage{cite}
\fi

\ifCLASSINFOpdf%
  \usepackage[pdftex]{graphicx}
  \graphicspath{{../pdf/}{../jpeg/}}
  \DeclareGraphicsExtensions{.pdf,.jpeg,.png}
\else
  \usepackage[dvips]{graphicx}
  \graphicspath{{../eps/}}
  \DeclareGraphicsExtensions{.eps}
\fi

\usepackage[acronym]{glossaries}
\usepackage{algorithmic}
\usepackage{amsmath,amssymb,amsfonts}
\usepackage{cite}
\usepackage{enumitem}
\usepackage{listings}
\usepackage{fix-cm}
\usepackage[shortcuts]{extdash}
\usepackage{textcomp}
\usepackage[dvipsnames]{xcolor}
\usepackage[shortcuts]{extdash}

\usepackage[english]{babel}

\usepackage{graphicx}
\usepackage{booktabs}
\usepackage[referable]{threeparttablex}
\usepackage{tabularx}
\usepackage{multirow}
\usepackage{colortbl}
\usepackage{tikz}
\usepackage{cancel}
\usepackage{arydshln}
\usepackage[free-standing-units,per-mode=repeated-symbol,binary-units=true,detect-weight=true,detect-family=true]{siunitx}[=v2]

\usepackage[hidelinks]{hyperref}
\usepackage[capitalise,nameinlink]{cleveref}
\usepackage[stretch=50,shrink=50]{microtype}

\DeclareSIUnit\squaredkm{\si{km.s^{2}}}
\DeclareSIUnit\devspersquaredkm{dev/\si{km^{2}}}
\DeclareSIUnit\core{core}
\DeclareSIUnit\request{req}
\DeclareSIUnit\cycle{cycle}
\DeclareSIUnit\bps{bps}
\DeclareSIUnit\Bps{Bps}
\DeclareSIUnit\terabps{T\bps}
\DeclareSIUnit\terabeps{Ti\Bps}
\DeclareSIUnit\gigabps{G\bps}
\DeclareSIUnit\megabps{M\bps}
\DeclareSIUnit\op{OP}
\DeclareSIUnit\ops{OPS}
\DeclareSIUnit\flop{FLOP}
\DeclareSIUnit\flops{FLOPS}
\DeclareSIUnit\teraops{TOPS}
\DeclareSIUnit\gate{GE}
\DeclareSIUnit\ge{GE}
\DeclareSIUnit\mhz{MHz}
\DeclareSIUnit\ghz{GHz}
\DeclareSIUnit\ipc{IPC}
\DeclareSIUnit\ips{IPS}
\DeclareSIUnit\bits{bits}
\DeclareSIUnit[number-unit-product = ]\percent{\%}

\definecolor{grey0}{HTML}{1B2B34}
\definecolor{grey1}{HTML}{343D46}
\definecolor{grey2}{HTML}{4F5B66}
\definecolor{grey3}{HTML}{65737E}
\definecolor{grey4}{HTML}{A7ADBA}
\definecolor{grey5}{HTML}{8CABA8}
\definecolor{grey6}{HTML}{BECBD2}
\definecolor{grey7}{HTML}{E9EEF0}
\definecolor{red}{HTML}{A8322D}
\definecolor{orange}{HTML}{FF8C00}
\definecolor{yellow}{HTML}{F8BD3F}
\definecolor{green}{HTML}{92BA51}
\definecolor{teal}{HTML}{108A7C}
\definecolor{blue}{HTML}{2A4857}
\definecolor{purple}{HTML}{875A78}
\definecolor{brown}{HTML}{805A4D}
\definecolor{color1}{HTML}{f94144}
\definecolor{color2}{HTML}{f9c74f}
\definecolor{color3}{HTML}{90be6d}
\definecolor{color4}{HTML}{43aa8b}
\definecolor{color5}{HTML}{577590}
\definecolor{PulpGreen}{HTML}{168638}
\definecolor{PulpBlue}{HTML}{1269b0}
\definecolor{PulpRed}{HTML}{a8322c}
\definecolor{PulpYellow}{HTML}{f2c100}

\PackageWarning{Citation missing:}{#1!}

\PackageWarning{TODO:}{#1!}

\newcommand\eg{e.g.,\ }

\newcommand\etal{et\penalty50\ al.\ }

\newcommand\terapool[1]{\ensuremath{\text{Tera\-Pool}_{#1}}}

\newcommand{\viewpoint}[1]{(\textbf{\romannumeral #1}.)}
\newcommand{\circled}[1]{%
    \tikz[baseline=(char.base)]{
        \node[shape=circle,draw,inner sep=0.8pt, thick] (char) {\footnotesize\textbf{#1}};}}

\ifx\showrevision\undefined
\else
    \AddToShipoutPictureFG{%
        \put(%
            8mm,%
            \paperheight-1.5cm%
            ){\vtop{{\null}\makebox[0pt][c]{%
                \rotatebox[origin=c]{90}{%
                    \huge\textcolor{red!75}{\reviewpass}%
                }%
            }}%
        }%
    }
    \AddToShipoutPictureFG{%
        \put(%
            \paperwidth-6mm,%
            \paperheight-1.5cm%
            ){\vtop{{\null}\makebox[0pt][c]{%
                \rotatebox[origin=c]{90}{%
                    \huge\textcolor{red!30}{ETH Z\"urich - Unpublished - Confidential - Draft - Copyright Yichao, Marco 2024}%
                }%
            }}%
        }%
    }
\fi

%% file: files/acronyms.tex
\newacronym{pe}{PE}{Processing Element}
\newacronym{cc}{CC}{Core-Complex}
\newacronym{ai}{AI}{Artificial Intelligence}
\newacronym{ml}{ML}{Machine Learning}
\newacronym{cnn}{CNN}{Convolutional Neural Network}
\newacronym{cpu}{CPU}{Central Processing Unit}
\newacronym{gpgpu}{GP-GPU}{General Purpose Graphics Processing Unit}
\newacronym{gpu}{GPU}{Graphics Processing Unit}
\newacronym{sm}{SM}{Streaming Multiprocessor}
\newacronym{asic}{ASIC}{Application Specific Integrated Circuit}
\newacronym{hpc}{HPC}{High Performance Computing}
\newacronym{npu}{NPU}{Neural Processing Unit}

\newacronym{5g}{5G}{5th-Generation}
\newacronym{6g}{6G}{6th-Generation}
\newacronym{pusch}{PUSCH}{Physical Uplink Shared Channel}
\newacronym{sdr}{SDR}{Software Defined Radio}
\newacronym{phy}{PHY}{Physical Layer}
\newacronym{ofdma}{OFDMA}{Orthogonal Frequency Division Multiple Access}
\newacronym{bf}{BF}{Beam Forming}
\newacronym{mimo}{MIMO}{Multiple-Input, Multiple-Output}
\newacronym{oran}{O-RAN}{Open Radio Access Networks}
\newacronym{ran}{RAN}{Radio Access Networks}
\newacronym{cots}{COTS}{Commercial-Off-The-Shelf}
\newacronym{dnn}{DNN}{Deep Neural Network}
\newacronym{llm}{LLM}{Large Language Model}

\newacronym{axpy}{AXPY}{A Times X Plus Y}
\newacronym{dotp}{DOTP}{Dot Product}
\newacronym{matmul}{MatMul}{Matrix Multiplication}
\newacronym{gemm}{GEMM}{General Matrix Multiplications}
\newacronym{spmmadd}{SpMMadd}{Sparse Matrix–Matrix Addition}
\newacronym{csr}{CSR}{Control and Status Register}
\newacronym{fft}{FFT}{Fast Fourier Transform}
\newacronym{che}{CHE}{Channel Estimation}
\newacronym{sysinv}{SysInv}{Linear System Inversion}
\newacronym{choldec}{CholDec}{Cholesky Decomposition}
\newacronym{ofdm}{OFDM}{Orthogonal Frequency Division Multiplexing}
\newacronym{tti}{TTI}{Transition Time Interval}

\newacronym{ppa}{PPA}{Power, Performance and Area}
\newacronym[longplural={Systems-on-Chip}]{soc}{SoC}{System-on-Chip}
\newacronym{numa}{NUMA}{Non-Uniform Memory Access}
\newacronym{isa}{ISA}{Instruction Set Architecture}
\newacronym{vliw}{VLIW}{Very Long Instruction Word}
\newacronym{alu}{ALU}{Arithmetic Logic Unit}
\newacronym{fp}{FP}{Floating Point}
\newacronym{fpu}{FPU}{Floating Point Unit}
\newacronym{ipu}{IPU}{Integer Processing Unit}
\newacronym{dsp}{DSP}{Digital Signal Processing}
\newacronym{simd}{SIMD}{Single Instruction Multiple Data}
\newacronym{spmd}{SPMD}{Single Program Multiple Data}
\newacronym{simt}{SIMT}{Single Instruction Multiple Threads}
\newacronym{eda}{EDA}{Electronic Design Automation}
\newacronym{ge}{GE}{Gate Equivalent}
\newacronym{fo4}{FO4}{Fan-Out-of-4}
\newacronym{IDol}{I\$}{Instruction Cache}
\newacronym{sram}{SRAM}{Static Random-Access Memory}
\newacronym{dram}{DRAM}{Dynamic Random-Access Memory}
\newacronym{spm}{SPM}{Scratchpad Memory}
\newacronym{tcdm}{TCDM}{Tightly Coupled Data Memory}
\newacronym{dma}{DMA}{Direct Memory Access}
\newacronym{axi}{AXI4}{Advanced eXtensible Interface}
\newacronym{noc}{NoC}{Network on Chip}
\newacronym{hbm}{HBM2E}{High Bandwidth Memory}
\newacronym{add}{add}{Add}
\newacronym{mul}{mul}{Multiply}
\newacronym{mac}{MAC}{Multiply Accumulate}
\newacronym{pmac}{p.mac}{Post-increment Multiply-accumulate}
\newacronym{amat}{AMAT}{Average Memory Access Time}
\newacronym{pmf}{PMF}{Probability Mass Function}
\newacronym{fifo}{FIFO}{First In First Out}
\newacronym{fpss}{FP-SS}{Floating Point Sub-System}
\newacronym{sdotp}{SDOTP}{Sum Dot Product}
\newacronym{scm}{SCM}{Standard Cell Memory}
\newacronym{hbml}{HBML}{High Bandwidth Memory Link}
\newacronym{ddr}{DDR}{Double Data Rate}
\newacronym{rtl}{RTL}{Register-Transfer Level}
\newacronym{divsqrt}{DIVSQRT}{Division and Square Root}
\newacronym{rvv}{RVV}{RISC-V Vector}

\newacronym{ipc}{IPC}{instructions-per-cycle}
\newacronym{wfi}{WFI}{wait-for-interrupt}
\newacronym{raw}{RAW}{read-after-write}
\newacronym{lsu}{LSU}{Load Store Unit}
\newacronym{ins}{INS}{instruction}
\newacronym{beol}{BEOL}{Back-End-of-Line}
\newacronym{pnr}{PnR}{Place and Route}
\newacronym{fc}{FC}{Fully-Connected}
\newacronym{soa}{SoA}{State-of-the-Art}
\newacronym{gsm}{GSM}{Graduate Student Member}
\newacronym{edp}{EDP}{Energy-Delay Product}

%% file: files/title_authorship.tex
\bstctlcite{IEEEexample:BSTcontrol}

\title{TeraPool: A Physical Design Aware, 1024 RISC-V Cores Shared-L1-Memory Scaled-up Cluster Design with High Bandwidth Main Memory Link}

\author{Yichao~Zhang, Marco~Bertuletti, Chi~Zhang, Samuel~Riedel, Diyou~Shen, Bowen~Wang, Alessandro~Vanelli-Coralli and~Luca~Benini
  \IEEEcompsocitemizethanks{%
    \IEEEcompsocthanksitem{} Yichao~Zhang, Marco~Bertuletti, Chi~Zhang, Samuel~Riedel, Diyou~Shen and Bowen~Wang are with the
     Integrated Systems Laboratory (IIS), ETH Zurich, 8092 Zurich, Switzerland. E-mail:
     \{yiczhang, mbertuletti, chizhang, sriedel, dishen, bowwang\}@iis.ee.ethz.ch
    \IEEEcompsocthanksitem{} Luca~Benini and Alessandro~Vanelli-Coralli are with the IIS, ETH Zurich, and also with the Department of Electrical, Electronic and Information Engineering (DEI), University of Bologna, 40126 Bologna, Italy. E-mail: \{lbenini, avanelli\}@iis.ee.ethz.ch.
  }
}

\markboth{IEEE TRANSACTIONS ON COMPUTERS,~Vol.~74, Issue.~11, November~2025}%
{Zhang \MakeLowercase{\etal}: TeraPool: A 1024 RISC-V Cores Shared-L1-Memory Cluster Design}

\newcommand\copyrighttext{\footnotesize \textcopyright This work has been submitted to the IEEE for possible publication. Copyright may be transferred without notice, after which this version may no longer be accessible.}
\newcommand\copyrightnotice{%
    \begin{tikzpicture}[remember picture,overlay]
        \node[anchor=south,yshift=10pt] at (current page.south) {\fbox{\parbox{\dimexpr\textwidth-\fboxsep-\fboxrule\relax}{\copyrighttext}}};
    \end{tikzpicture}%
    \vspace{-4em}
}

%% file: tables/congestion.tex
\begin{threeparttable}
\begin{tabular}{c|ccc|c|c}
\hline
\multirow{2}{*}{\begin{tabular}[c]{@{}c@{}}Interco.\\ Complexity\end{tabular}} & \multicolumn{3}{c|}{Congestion\textsuperscript{*}} & \multirow{2}{*}{\begin{tabular}[c]{@{}c@{}}Area\\ (kGE)\end{tabular}} & \multirow{2}{*}{\begin{tabular}[c]{@{}c@{}}Critical\\ Path (ns)\end{tabular}} \\ \cline{2-4}
                                     & H         & V        & Overall  &                                                                       &                                                                                \\ \hline
256                                  & 0.13\%    & 0.07\%   & 0.10\%   & 109                                                               & 0.59                                                                          \\
512                                  & 0.26\%    & 0.11\%   & 0.19\%   & 196                                                               & 0.73                                                                          \\
1024                                 & 0.56\%    & 0.12\%   & 0.34\%   & 361                                                               & 0.91                                                                          \\
1280                                 & 1.72\%    & 0.47\%   & 1.09\%   & 503                                                               & 1.06                                                                          \\
1536                                 & 3.25\%    & 0.82\%   & 2.04\%   & 669                                                               & 1.08                                                                          \\
2048                                 & 34.46\%   & 15.09\%  & 24.77\%  & 923                                                               & 1.13                                                                          \\
3072                                 & 172.30\%  & 294.31\% & 233.31\% & 1274                                                              & 1.27                                                                          \\
4096                                 & 247.10\%  & 368.90\% & 308.00\% & 1485                                                              & 1.47                                                                          \\ \hline
\end{tabular}
\end{threeparttable}

%% file: tables/interconnection.tex
\begin{threeparttable}
\begin{tabular}{c|cccccc|cccc}
\hline
                                                                                              & \multicolumn{6}{c|}{\cellcolor[HTML]{9FC5E8}Interconnect Quality}                                                                                                                                                                                                                                                                                                                       & \multicolumn{4}{c}{\cellcolor[HTML]{EA9999}Design Challenge}                                                                                                                                                                                      \\
\multirow{-2}{*}{\begin{tabular}[c]{@{}c@{}}Hierarchy\\ Interco.\textsuperscript{*}\end{tabular}} & \begin{tabular}[c]{@{}c@{}}ZeroLd \\ (cyc)\end{tabular} & \begin{tabular}[c]{@{}c@{}}AMAT \\ (cyc)\end{tabular} & \begin{tabular}[c]{@{}c@{}}Throughput\\ (req/pe/cyc)\end{tabular} & \begin{tabular}[c]{@{}c@{}}Total\\ Complex.\end{tabular} & \begin{tabular}[c]{@{}c@{}}Critical\\ Complex.\end{tabular} & \begin{tabular}[c]{@{}c@{}}Critical\\ Comb.\\ Delay\end{tabular} & \begin{tabular}[c]{@{}c@{}}Physical\\ Routing\end{tabular} & \begin{tabular}[c]{@{}c@{}}Path \\ Balance\end{tabular} & \begin{tabular}[c]{@{}c@{}}Base\\ Block\\ Scale\end{tabular} & \begin{tabular}[c]{@{}c@{}}Interco.\\ Perform.\end{tabular} \\ \hline
1024C                                                                                         & 1.000                                                   & 1.130                                                 & 0.885                                                                        & 4194304                                                  & {\color[HTML]{9A0000} 4194304}                              & {\color[HTML]{9A0000} 22}                                        & \cellcolor[HTML]{EA9999}                                   & \cellcolor[HTML]{D9EAD3}                                & \cellcolor[HTML]{EA9999}                                     & \cellcolor[HTML]{D9EAD3}                                    \\
4C-256T                                                                                       & 2.992                                                   & 6.081                                                 & 0.245                                                                        & 87040                                                    & {\color[HTML]{9A0000} 65536}                                & 16                                                               & \cellcolor[HTML]{EA9999}                                   & \cellcolor[HTML]{D9EAD3}                                & \cellcolor[HTML]{EA9999}                                     & \cellcolor[HTML]{D9EAD3}                                    \\
8C-128T                                                                                       & 2.984                                                   & {\color[HTML]{9A0000} 10.075}                         & {\color[HTML]{9A0000} 0.124}                                                 & 54272                                                    & {\color[HTML]{9A0000} 16384}                                & 14                                                               & \cellcolor[HTML]{EA9999}                                   & \cellcolor[HTML]{D9EAD3}                                & \cellcolor[HTML]{EA9999}                                     & \cellcolor[HTML]{EA9999}                                    \\
16C-64T                                                                                       & 2.969                                                   & {\color[HTML]{9A0000} 18.077}                         & {\color[HTML]{9A0000} 0.062}                                                 & 74752                                                    & {\color[HTML]{9A0000} 4096}                                 & 12                                                               & \cellcolor[HTML]{EA9999}                                   & \cellcolor[HTML]{D9EAD3}                                & \cellcolor[HTML]{EA9999}                                     & \cellcolor[HTML]{EA9999}                                    \\
4C-16T-16G                                                                                    & 4.867                                                   & 5.318                                                 & 0.431                                                                        & 163840                                                   & 320                                                         & 8.3                                                              & \cellcolor[HTML]{D9EAD3}                                   & \cellcolor[HTML]{EA9999}                                & \cellcolor[HTML]{D9EAD3}                                     & \cellcolor[HTML]{D9EAD3}                                    \\
4C-32T-8G                                                                                     & 4.742                                                   & 5.443                                                 & 0.409                                                                        & 122880                                                   & 1024                                                        & 10                                                               & \cellcolor[HTML]{D9EAD3}                                   & \cellcolor[HTML]{EA9999}                                & \cellcolor[HTML]{EA9999}                                     & \cellcolor[HTML]{D9EAD3}                                    \\
8C-16T-8G                                                                                     & 4.734                                                   & 5.794                                                 & 0.358                                                                        & 90112                                                    & 512                                                         & 9                                                                & \cellcolor[HTML]{D9EAD3}                                   & \cellcolor[HTML]{EA9999}                                & \cellcolor[HTML]{EA9999}                                     & \cellcolor[HTML]{D9EAD3}                                    \\
8C-32T-4G                                                                                     & 4.484                                                   & 6.676                                                 & 0.272                                                                        & 69632                                                    & 1024                                                        & 10                                                               & \cellcolor[HTML]{D9EAD3}                                   & \cellcolor[HTML]{D9EAD3}                                & \cellcolor[HTML]{EA9999}                                     & \cellcolor[HTML]{D9EAD3}                                    \\
16C-8T-8G                                                                                     & 4.719                                                   & 6.669                                                 & 0.273                                                                        & 110592                                                   & 1536                                                        & 10.6                                                             & \cellcolor[HTML]{D9EAD3}                                   & \cellcolor[HTML]{EA9999}                                & \cellcolor[HTML]{EA9999}                                     & \cellcolor[HTML]{D9EAD3}                                    \\
16C-16T-4G                                                                                    & 4.469                                                   & 8.612                                                 & {\color[HTML]{9A0000} 0.178}                                                 & 90112                                                    & 1280                                                        & 10.3                                                             & \cellcolor[HTML]{D9EAD3}                                   & \cellcolor[HTML]{D9EAD3}                                & \cellcolor[HTML]{EA9999}                                     & \cellcolor[HTML]{EA9999}                                    \\
4C-16T-4SG-4G                                                                                 & 6.367                                                   & 8.457                                                 & 0.270                                                                        & 121856                                                   & {\color[HTML]{9A0000} 4096}                                 & 12                                                               & \cellcolor[HTML]{EA9999}                                   & \cellcolor[HTML]{D9EAD3}                                & \cellcolor[HTML]{D9EAD3}                                     & \cellcolor[HTML]{D9EAD3}                                    \\
\rowcolor[HTML]{D9EAD3} 
8C-8T-4SG-4G                                                                                  & 6.359                                                   & 9.198                                                 & 0.230                                                                        & 89088                                                    & 1024                                                        & 10                                                               & \cellcolor[HTML]{D9EAD3}                                   & \cellcolor[HTML]{D9EAD3}                                & \cellcolor[HTML]{D9EAD3}                                     & \cellcolor[HTML]{D9EAD3}                                    \\
16C-4T-4SG-4G                                                                                 & 6.344                                                   & {\color[HTML]{9A0000} 11.049}                         & {\color[HTML]{9A0000} 0.159}                                                 & 109568                                                   & 1536                                                        & 10.6                                                             & \cellcolor[HTML]{D9EAD3}                                   & \cellcolor[HTML]{D9EAD3}                                & \cellcolor[HTML]{D9EAD3}                                     & \cellcolor[HTML]{EA9999}                                   \\ \hline
\end{tabular}
\end{threeparttable}

%% file: tables/SoA.tex
\setlength{\tabcolsep}{1.2pt} 
\begin{threeparttable}[]
\begin{tabular}{cc|ccc|ccccccc|c}
\cline{2-13}
\multicolumn{1}{l}{}                                                                           &\multicolumn{1}{l|}{}                            & \begin{tabular}[c]{@{}c@{}}Processing \\ Element (PE)\end{tabular} & Data Types                                                                                                                   & \begin{tabular}[c]{@{}c@{}}Execution\\ Model\end{tabular} & \begin{tabular}[c]{@{}c@{}}No.PEs\\ /Cluster\end{tabular} & \begin{tabular}[c]{@{}c@{}}Total \\ PEs\end{tabular} & \begin{tabular}[c]{@{}c@{}}Shared-L1\\ (MiB)\end{tabular} & \begin{tabular}[c]{@{}c@{}}L1/L2 Interco. \\ Bandwidth\\ (Byte/Cycle)\end{tabular} & \begin{tabular}[c]{@{}c@{}}L1 Latency\\ (cycles)\end{tabular}              & \begin{tabular}[c]{@{}c@{}}Peak (FL)OP\\ /cyc./Cluster$^{a}$\end{tabular}           & \begin{tabular}[c]{@{}c@{}}Scaling\\ Topology\end{tabular}                                    & \begin{tabular}[c]{@{}c@{}}Open\\ Source\end{tabular} \\ \cline{2-13}

\multirow{2}{*}{\raisebox{-8em}{\rotatebox{90}{\begin{tabular}[c]{@{}c@{}}\textbf{Multi-core}\\ (No.PEs \textless 100)\end{tabular}}}}   &\textbf{This work}                                                       & 32bit RISC-V                                                       & {\begin{tabular}[c]{@{}c@{}}Int32/16-SIMD;\\ FP32/16-SIMD;\\ Complex32\\ (16b Real\&Imag)\end{tabular}} & {\color[HTML]{009901} \textbf{SPMD}}                      & {\color[HTML]{009901} \textbf{1024}} & 1024                     & {\color[HTML]{009901} \textbf{4}}                         & {\color[HTML]{009901} \textbf{4096/1024}}                                          & {\color[HTML]{009901} \textbf{\begin{tabular}[c]{@{}c@{}}1-11\\ (NUMA)\end{tabular}}} & {\color[HTML]{009901} \textbf{2048}} & {\color[HTML]{009901} \textbf{\begin{tabular}[c]{@{}c@{}}Scaling-up\\ Crossbar\end{tabular}}} & {\color[HTML]{009901} \textbf{Yes}}                   \\ \cline{2-13}

& \textbf{\begin{tabular}[c]{@{}c@{}}Kalray \\ MPPA3-80~\cite{Kalery_2021}\end{tabular}}      & 64bit VLIW                                                         & \begin{tabular}[c]{@{}c@{}}Int16/8;\\ FP64/32/16\end{tabular}                                                                & \begin{tabular}[c]{@{}c@{}}SPMD;\\ LWI\end{tabular}       & 16  & 80                                                      & 3.8                                                       & 64/1024                                                                            & 23                                                                                  & 32  & \begin{tabular}[c]{@{}c@{}}Scaling-out\\ 2D-mesh NoC\end{tabular}                             & No                                                    \\

& \textbf{\begin{tabular}[c]{@{}c@{}}Ramon\\ RC64~\cite{Ramon_2016, Ramon_2021}\end{tabular}}            & 32bit VLIW                                                         & Int32/16/8                                                                                                                   & MIMD                                                      & 64    & 64                                                    & 3.8                                                       & 1024/N.A.                                                                          & N.A.                                                                    & 128              & \begin{tabular}[c]{@{}c@{}}Scaling-up\\ Crossbar\end{tabular}                                 & No                                                    \\ \cline{2-13}

& \textbf{\begin{tabular}[c]{@{}c@{}}TensTorrent~\cite{Tenstorrent_2021} \\ Wormhole\end{tabular}} & 32bit RISC-V                                                       & FP32/16/8                                                                                                                    & SIMD                                                      & 5  & 400                                                       & 1.43                                                      & 20/N.A.                                                                            & \textgreater~4$^{b}$                                                    & 10                    & \begin{tabular}[c]{@{}c@{}}Scaling-out\\ 2D-mesh NoC\end{tabular}                             & No                                                    \\

\multirow{5}{*}{\rotatebox{90}{\begin{tabular}[c]{@{}c@{}}\textbf{Many-core}\\ (No.PEs \textgreater 100)\end{tabular}}} & \textbf{\begin{tabular}[c]{@{}c@{}}Esperanto\\ ET-SoC-1~\cite{ET_soc_2022}\end{tabular}}    & 64bit RVV                                                          & \begin{tabular}[c]{@{}c@{}}Int8/N.A.;\\ FP32/16\end{tabular}                                                                 & SIMD                                                      & 32   & 1088                                                     & 3.8                                                       & 256/32                                                                             & N.A.                                                                    & 64              & \begin{tabular}[c]{@{}c@{}}Scaling-out\\ 2D-mesh NoC\end{tabular}                             & No                                                    \\

& \textbf{\begin{tabular}[c]{@{}c@{}}Nvidia\\ H100~\cite{nvidia_h100_2023}\end{tabular}}           & 64/32bit PTX                                                       & \begin{tabular}[c]{@{}c@{}}Int8; FP64/16/8;\\ BF16;TF32\end{tabular}                                                         & SIMT                                                      & \begin{tabular}[c]{@{}c@{}}128\\ (SM)\end{tabular} & 18432   & \begin{tabular}[c]{@{}c@{}}0.244\\ (SM)\end{tabular}  & 128/N.A.                                                                           & \begin{tabular}[c]{@{}c@{}}16.57$^{c}$\\ (Average)\end{tabular}            & 1736                 & \begin{tabular}[c]{@{}c@{}}Scaling-out\\ Data-Driven\end{tabular}                             & No                                                    \\ \cdashline{2-13}

& \textbf{\begin{tabular}[c]{@{}c@{}}Hammer$^{d}$\\Blade~\cite{Hammerblade_2024}\end{tabular}}                                                          & 32bit RISC-V                                                       & \begin{tabular}[c]{@{}c@{}}Int32/-;\\ FP32/-\end{tabular}                                                & SPMD                                                      & \begin{tabular}[c]{@{}c@{}}128\\ (Cell)\end{tabular}  & 2048                                                       & \begin{tabular}[c]{@{}c@{}}0.5\\ (Cell)\end{tabular}                                                     & 512/N.A.                                                                             & \begin{tabular}[c]{@{}c@{}}4 +\\2 x hops\end{tabular}                                                                             & 256        & \begin{tabular}[c]{@{}c@{}}Scaling-out\\ 2D-ruche NoC\end{tabular}                                & \color[HTML]{009901} \textbf{Yes}                                                   \\

& \textbf{Occamy~\cite{Occamy_2025}}                                                          & 64bit RISC-V                                                       & \begin{tabular}[c]{@{}c@{}}Int64/32/16-SIMD;\\ FP64/32/16/8-SIMD\end{tabular}                                                & SPMD                                                      & 8  & 432                                                       & 0.125                                                     & 32/256                                                                             & 1                                                                       & 32              & \begin{tabular}[c]{@{}c@{}}Scaling-out\\ Crossbar\end{tabular}                                & \color[HTML]{009901} \textbf{Yes}                                                   \\

& \textbf{MemPool~\cite{Mempool_2023}}                                                         & 32bit RISC-V                                                       & Int32/16-SIMD                                                                                                                & SPMD                                                      & 256     & 256                                                  & 1                                                         & 1024/256                                                                           & \begin{tabular}[c]{@{}c@{}}1-5\\ (NUMA)\end{tabular}                    & 512              & \begin{tabular}[c]{@{}c@{}}Scaling-up\\ Crossbar\end{tabular}                                 & \color[HTML]{009901} \textbf{Yes}                                                   \\ \cline{2-13}

\end{tabular}
\end{threeparttable}

%% file: tables/transfer_vs_IPC.tex
\begin{threeparttable}
\begin{tabular}{c|c|cc|cc}
\hline
                                                                                 &                                                                                            & \multicolumn{2}{c|}{\begin{tabular}[c]{@{}c@{}}Local Data Distrib. \\ Kernel (AXPY f32)\end{tabular}}                             & \multicolumn{2}{c}{\begin{tabular}[c]{@{}c@{}}Global Data Distrib.$^{**}$\\ Kernel (MatMul f32)\end{tabular}}                            \\ \cline{3-6} 
\multirow{-2}{*}{}                                                               & \multirow{-2}{*}{\begin{tabular}[c]{@{}c@{}}Cluster's\\ Max. Tiling \\ (MiB)\end{tabular}} & \begin{tabular}[c]{@{}c@{}}Main Mem.\\ Transfer \\ Byte/FLOP\end{tabular} & \begin{tabular}[c]{@{}c@{}}Comp.\\ IPC\end{tabular} & \begin{tabular}[c]{@{}c@{}}Main Mem.\\ Transfer \\ Byte/FLOP\end{tabular} & \begin{tabular}[c]{@{}c@{}}Comp.\\ IPC\end{tabular} \\ \hline
\begin{tabular}[c]{@{}c@{}}Scaled-up\\ Ultra-Large Cluster\\ (\textbf{This work}$^{*}$)\end{tabular} & {\color[HTML]{009901} \textbf{4.00}}                                                       & {\color[HTML]{009901} \textbf{6.00}}                                        & {\color[HTML]{009901} \textbf{0.85}}                & {\color[HTML]{009901} \textbf{0.009}}                                       & 0.70                                                \\ \hline
\begin{tabular}[c]{@{}c@{}}Scaled-up\\ Large Cluster\\ (\textbf{MemPool~\cite{Mempool_2023}}$^{*}$)\end{tabular}   & 1.00                                                                                       & 6.00                                                                        & 0.85                                                & 0.016                                                                       & 0.88                                                \\ \hline
\begin{tabular}[c]{@{}c@{}}Scaled-out\\ Many Clusters\\ (\textbf{Occamy~\cite{Occamy_2025}}$^{*}$)\end{tabular}        & 0.125                                                                                       & 6.00                                                                        & 0.85                                                & 0.062                                                                       & 0.89                                                \\ \hline
\end{tabular}
\end{threeparttable}

%% file: files/biography.tex
\vspace{-15mm}
\begin{IEEEbiography}[{\includegraphics[width=1in,height=1.25in,keepaspectratio,clip]{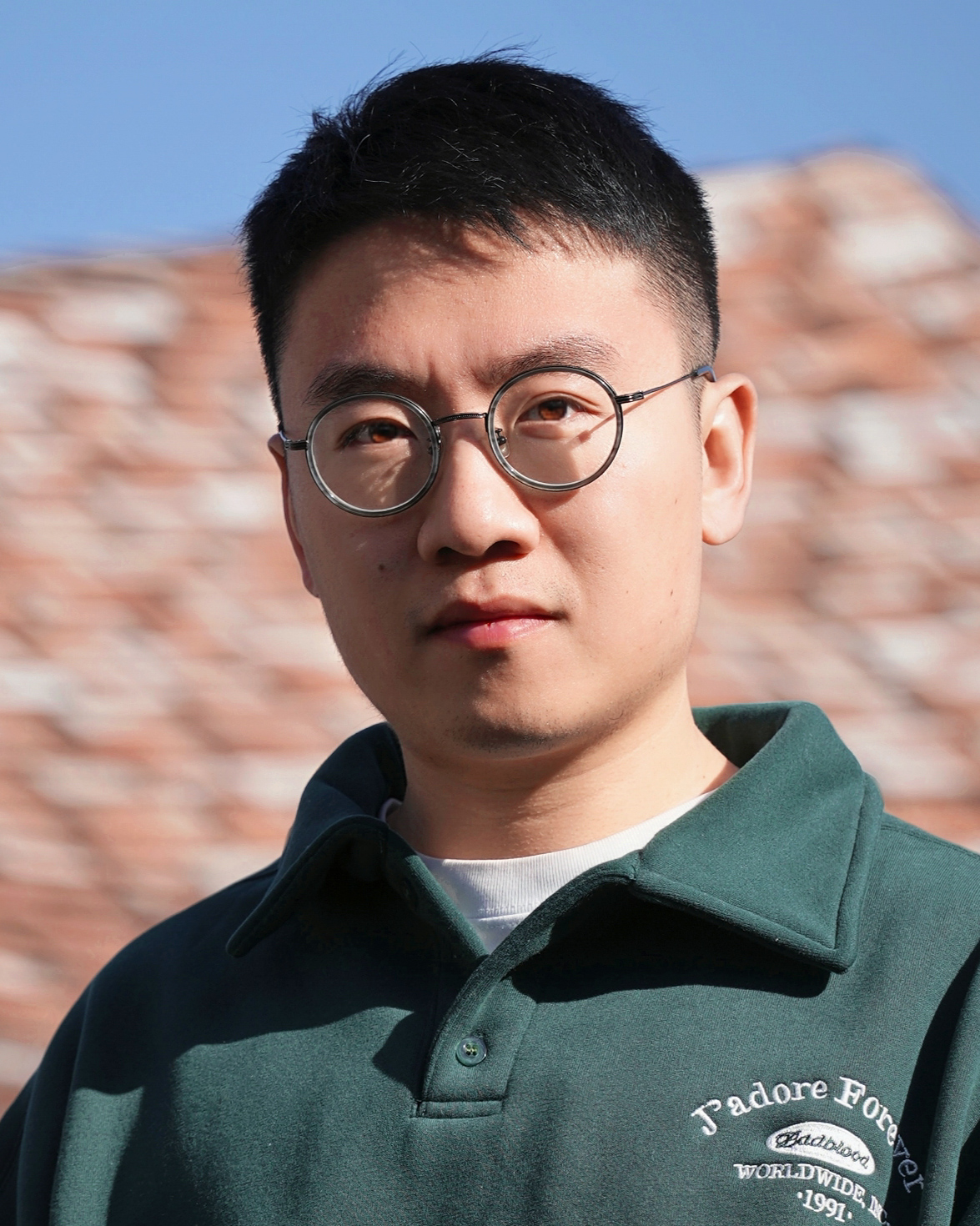}}]{Yichao Zhang}
received his B.Eng. degree from Heilongjiang University China in 2015 and his M.Sc. degree from Nanyang Technological University Singapore in 2017. He served in the physical VLSI design at Cadence Design Systems and MediaTek Singapore until 2021. He is currently pursuing a Ph.D. degree in the Digital Circuits and Systems group of Prof. Benini. His research interests include physically feasible, many-core parallel computing architectures, and SIMD processing.
\end{IEEEbiography}

\vspace{-5mm}
\begin{IEEEbiography}[{\includegraphics[width=1in,height=1.25in,keepaspectratio,clip]{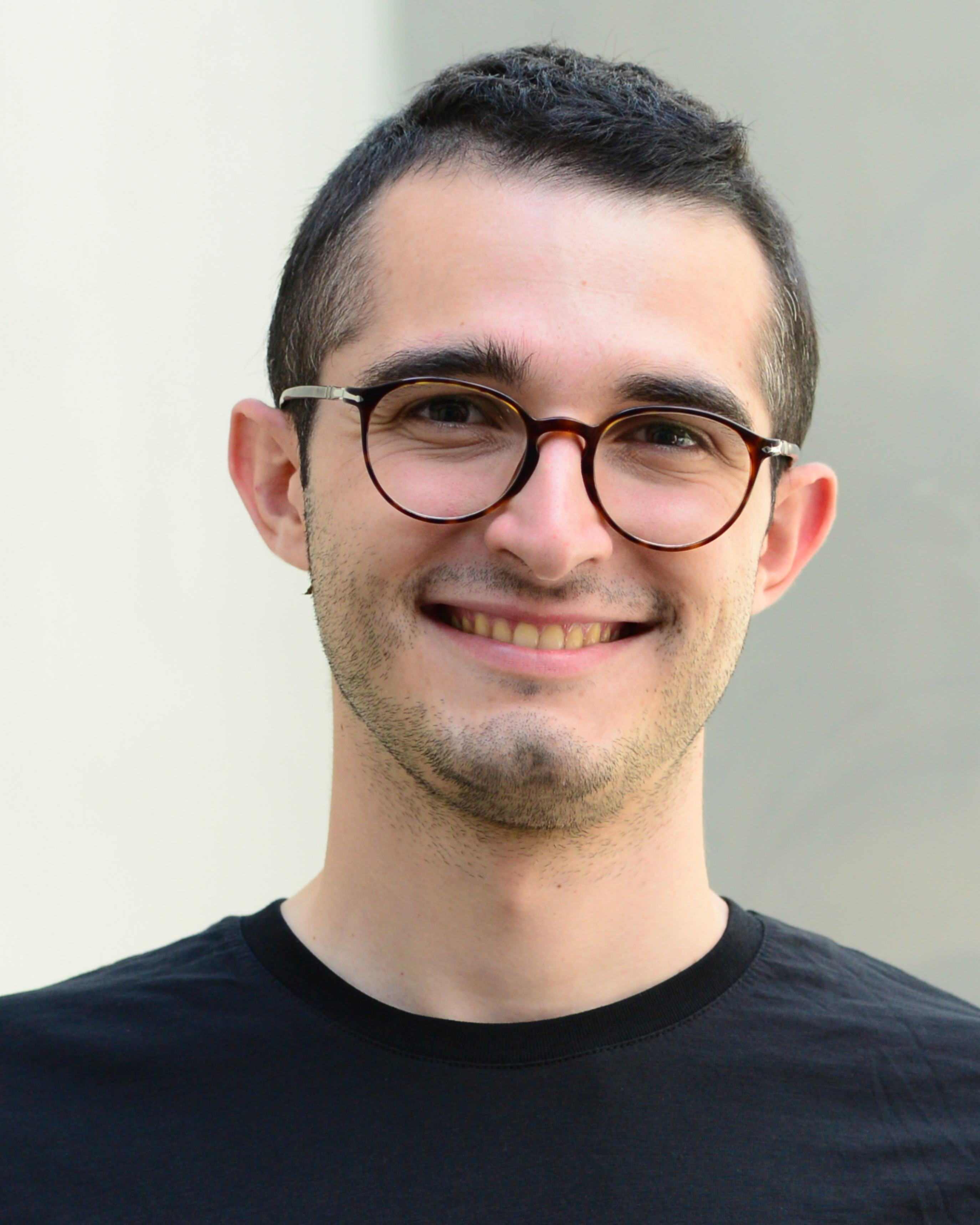}}]{Marco Bertuletti}
received his B.Sc. and M.Sc. degrees in Electrical Engineering at Politecnico di Milano, Milano, Italy. He is currently pursuing a Ph.D. degree in the Digital Circuits and Systems group of Prof. Benini. His research interests include the design of multi and Many-Core clusters of RISC-V processors for next-generation telecommunications.
\end{IEEEbiography}

\vspace{-5mm}
\begin{IEEEbiography}[{\includegraphics[width=1in,height=1.25in,keepaspectratio,clip]{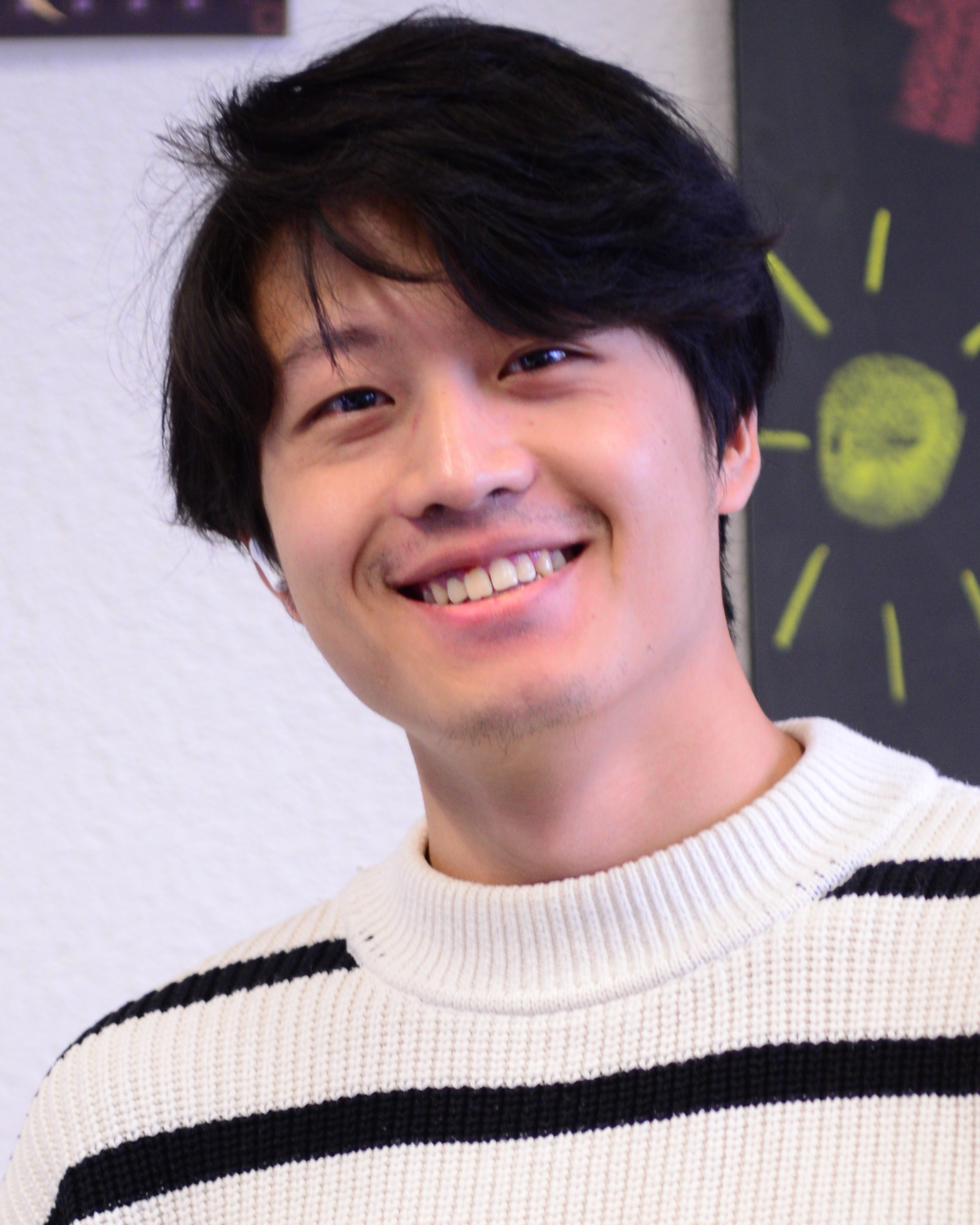}}]{Chi Zhang}
received his B.Sc. degree from Huazhong University of Science and Technology China in 2019 and his M.Sc. degree from KTH Royal Institute of Technology Sweden in 2022. He is currently pursuing a Ph.D. degree in the Digital Circuits and Systems group of Prof. Benini. His research interests include high-performance computing, memory systems, and near-memory computing.
\end{IEEEbiography}

\vspace{-5mm}
\begin{IEEEbiography}[{\includegraphics[width=1in,height=1.25in,keepaspectratio,clip]{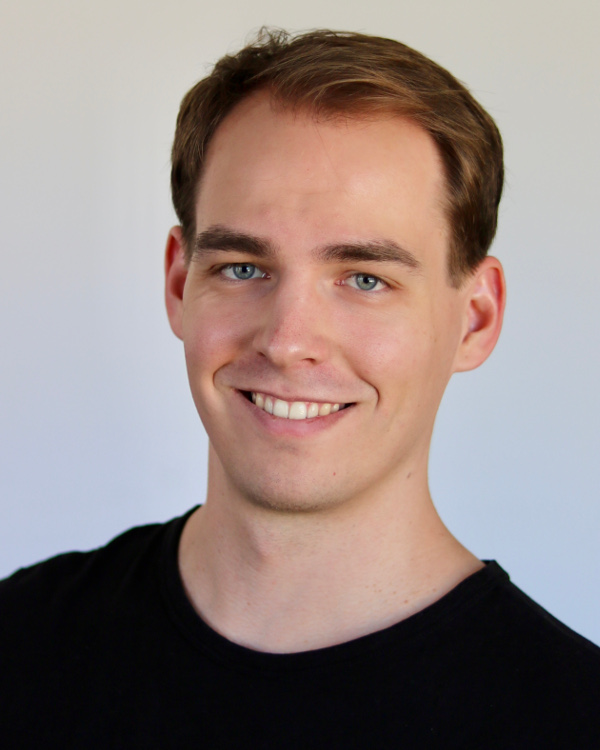}}]{Samuel Riedel}
received his B.Sc. and M.Sc. degrees in Electrical Engineering and Information Technology at ETH Zurich in 2017 and 2019, respectively. He is currently pursuing a Ph.D. degree in the Digital Circuits and Systems group of Prof. Benini. His research interests include computer architecture, focusing on manycore systems and their programming model.
\end{IEEEbiography}

\vspace{-5mm}
\begin{IEEEbiography}[{\includegraphics[width=1in,height=1.25in,keepaspectratio,clip]{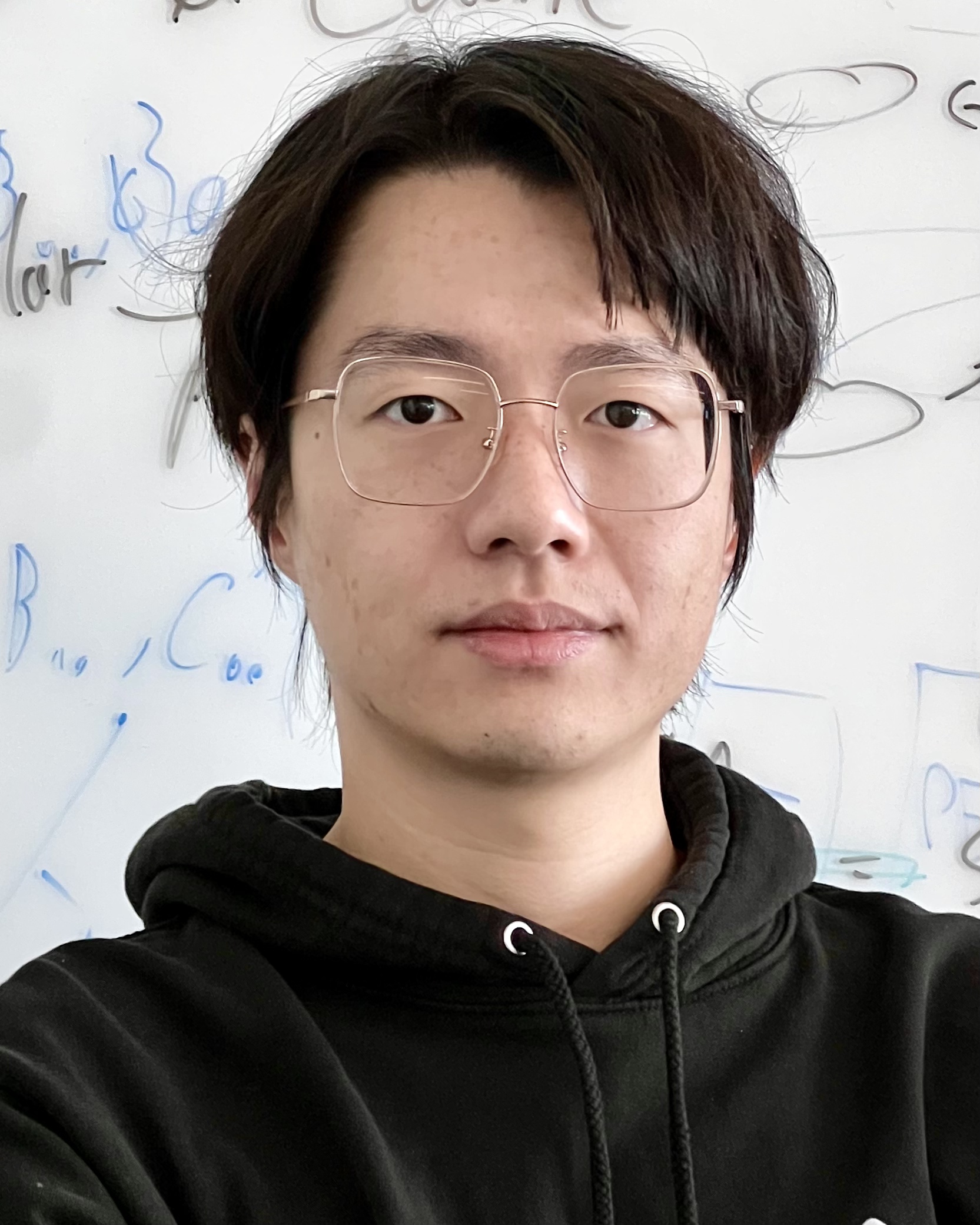}}]{Diyou Shen}
received his B.Sc. degree from University of Colorado Boulder in 2021 and his M.Sc. degree in Electrical Engineering and Information Technology from ETH Zurich in 2024. He is currently pursuing a Ph.D. degree in the Digital Circuits and Systems group of Prof. Benini. His research interests include high-performance computing and fault-tolerance interconnection.
\end{IEEEbiography}

\vspace{-5mm}
\begin{IEEEbiography}[{\includegraphics[width=1in,height=1.25in,keepaspectratio,clip]{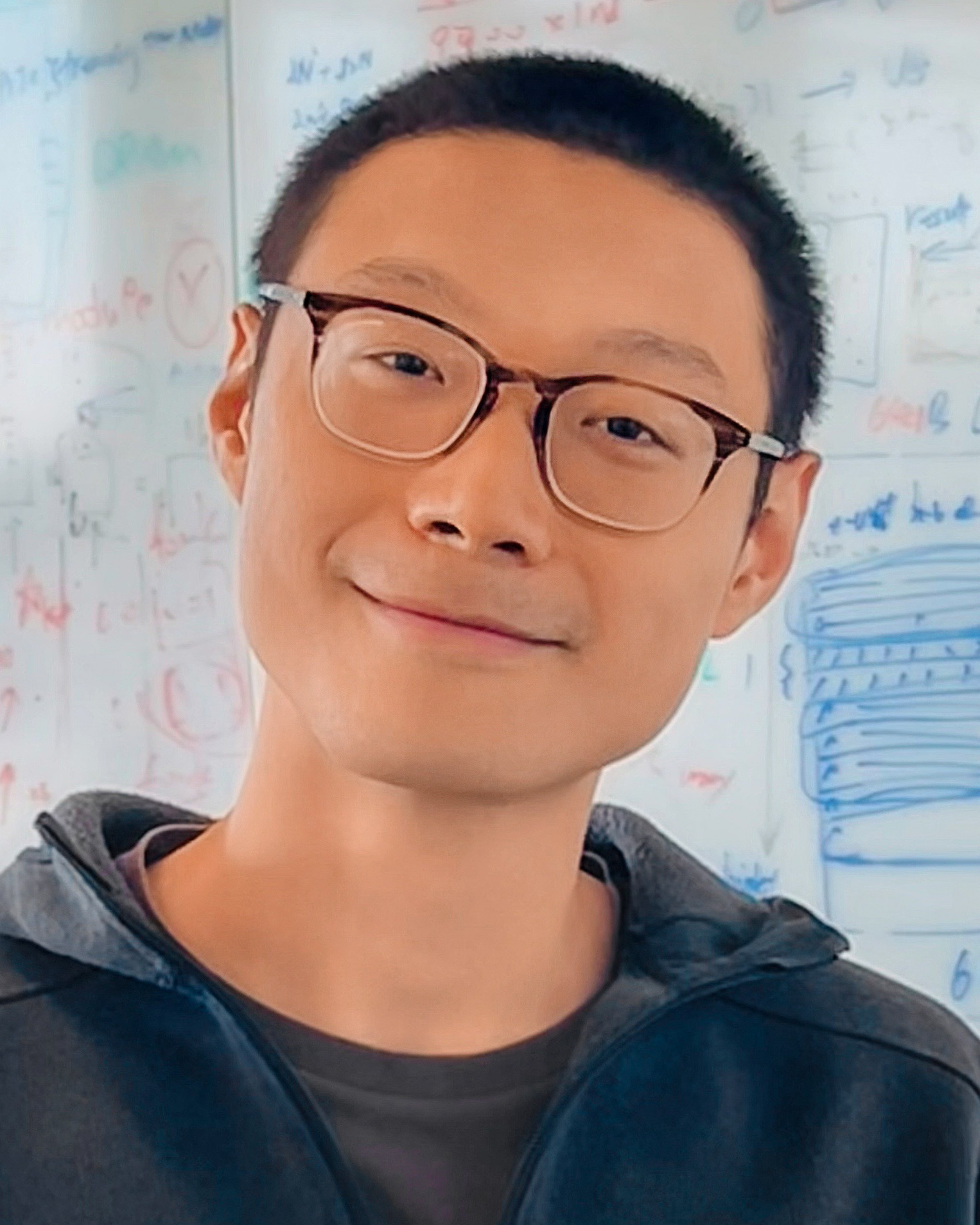}}]{Bowen Wang}
received his B.Sc. degree from Harbin Institute of Technology in 2021 and M.Sc. degrees in Electrical Engineering and Information Technology at ETH Zurich in 2024. He is currently pursuing a Ph.D. degree in the Digital Circuits and Systems group of Prof. Benini. His research interests include computer architecture, flexible manycore systems and neural network deployment.
\end{IEEEbiography}

\vspace{-5mm}
\begin{IEEEbiography}
[{\includegraphics[width=1in,height=1.25in,keepaspectratio,clip]{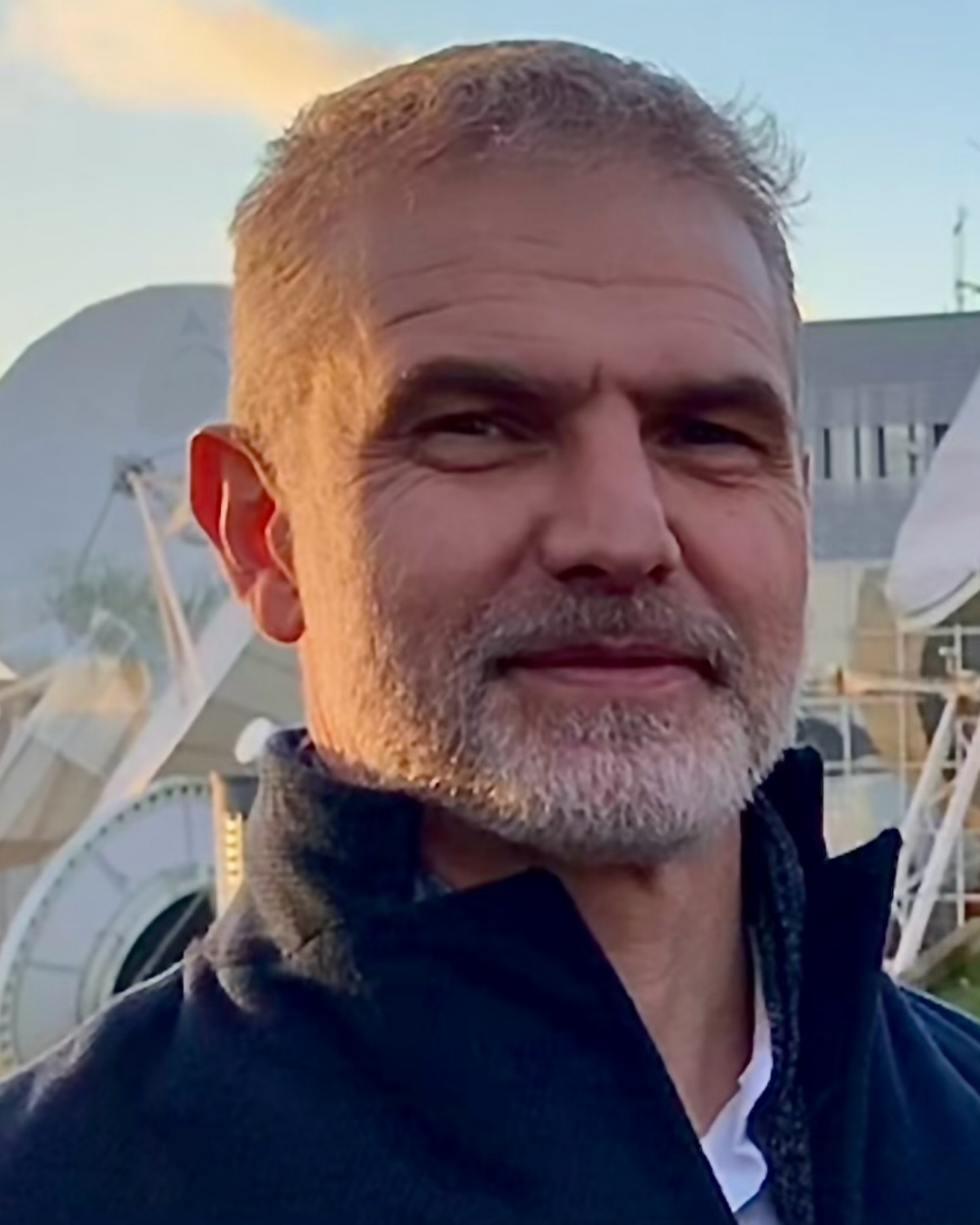}}]{Alessandro Vanelli-Coralli}
is Full Professor at the Università di Bologna and is a Senior Scientist at ETH Z\"urich. His research activity focuses on Wireless Communication with specific emphasis on Satellite Communications. He participates in national and international research projects on satellite mobile communication systems and Prime Contractor for several European Space Agency and European Commission funded projects. Dr. Vanelli-Coralli has been the general co-chairman of the IEEE ASMS Conference since 2010.
\end{IEEEbiography}

\vspace{-5mm}
\begin{IEEEbiography}[{\includegraphics[width=1in,height=1.25in,keepaspectratio,clip]{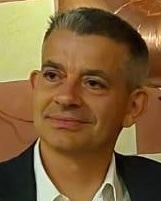}}]{Luca Benini}
holds the chair of digital Circuits and systems at ETHZ and is Full Professor at the Università di Bologna. He received a PhD from Stanford University. His research interests are in energy-efficient parallel computing systems, smart sensing micro-systems and machine learning hardware. He is a fellow of the ACM, a member of the Academia Europaea and of the Italian Academy of Engineering and Technology.
\end{IEEEbiography}

%% file: bib/terapool.bib
@IEEEtranBSTCTL{IEEEexample:BSTcontrol,
  CTLuse_forced_etal = "yes",
  CTLmax_names_forced_etal = "6",
  CTLnames_show_etal = "1",
  CTLuse_url = "yes",
  CTLdash_repeated_names = "yes"
}

@INPROCEEDINGS{Compute_Trend_2022,
   author = {Jaime Sevilla and Lennart Heim and Anson Ho and Tamay Besiroglu and Marius Hobbhahn and Pablo Villalobos},
   doi = {10.1109/IJCNN55064.2022.9891914},
   booktitle = {Int. Jt. Conf. Neural Netw.},
   title = {{Compute Trends Across Three Eras of Machine Learning}},
   organization = {IEEE},
   address = {Padua, Italy},
   pages = {1-8},
   year = {2022},
   month = {July}
}

@article{LLM_Survey_2_2024,
  author = {Zhu, Xunyu and Li, Jian and Liu, Yong and Ma, Can and Wang, Weiping},
  title = {{A Survey on Model Compression for Large Language Models}},
  journal = {Trans. Assoc. Comput. Linguist.},
  volume = {12},
  number = {1},
  pages = {1556-1577},
  year = {2024},
  month = {Nov.},
  doi = {10.1162/tacl_a_00704},
}

@ARTICLE{Survey_Hardware_Acc_2022,
  title = {{A survey on hardware accelerators: Taxonomy, Trends, Challenges, and Perspectives}},
  journal = {J. Syst. Archit.},
  volume = {129},
  note = {Article number 102561},
  pages = {1-51},
  month = {Aug.},
  year = {2022},
  issn = {1383-7621},
  doi = {https://doi.org/10.1016/j.sysarc.2022.102561},
  author = {Biagio Peccerillo and Mirco Mannino and Andrea Mondelli and Sandro Bartolini},
}

@MISC{50Years_trend_2022,
  author          = {Rupp, Karl},
  title           = {{50 Years of Microprocessor Trend Data}},
  year            = {2022},
  howpublished    = {GitHub Repo.},
  note            = "[Online]. Available: \url{https://github.com/karlrupp/microprocessor-trend-data}"
}

@ARTICLE{Near_Mem_2019,
title = {{Near-memory Computing: Past, Present, and Future}},
journal = {Microprocess. Microsyst.},
volume = {71},
note = {Article number 102868},
pages = {1-15},
year = {2019},
month = {Aug.},
issn = {0141-9331},
doi = {https://doi.org/10.1016/j.micpro.2019.102868},
author = {Gagandeep Singh and Lorenzo Chelini and Stefano Corda and Ahsan Javed Awan and Sander Stuijk and Roel Jordans and Henk Corporaal and Albert-Jan Boonstra},
}

@ARTICLE{Energy_Efficient_2022,
    author = {Muralidhar, Rajeev and Borovica-Gajic, Renata and Buyya, Rajkumar},
    title = {{Energy Efficient Computing Systems: Architectures, Abstractions and Modeling to Techniques and Standards}},
    year = {2022},
    volume = {54},
    number = {11},
    issn = {0360-0300},
    doi = {10.1145/3511094},
    journal = {ACM Comput. Surv.},
    month = {sep},
    articleno = {236},
    pages = {1-37},
}

@INPROCEEDINGS{Energy_problem_2014,
  author={M. Horowitz},
  booktitle={IEEE Int. Solid-State Circuits Conf. Dig. Tech. Pap.},
  title={{Computing’s Energy Problem (and What We Can Do About It)}},
  year={2014},
  month = {Feb.},
  organization = {IEEE},
  address = {San Francisco, CA, USA},
  pages={10-14},
  doi={10.1109/ISSCC.2014.6757323}
}

@ARTICLE{mppa256_2017,
  author={Ishii, Masahiro and Detrey, Jérémie and Gaudry, Pierrick and Inomata, Atsuo and Fujikawa, Kazutoshi},
  journal={IEEE Trans. Comput.}, 
  title={{Fast Modular Arithmetic on the Kalray MPPA-256 Processor for an Energy-Efficient Implementation of ECM}}, 
  year={2017},
  month = {May},
  volume={66},
  number={12},
  pages={2019-2030},
  doi={10.1109/TC.2017.2704082}
}

@ARTICLE{ET_soc_2022,
  author={Ditzel, David R. and others},
  journal={IEEE Micro}, 
  title={{Accelerating ML Recommendation With Over 1,000 RISC-V/Tensor Processors on Esperanto's ET-SoC-1 Chip}}, 
  year={2022},
  month = {Jan.},
  volume={42},
  number={3},
  pages={31-38},
  doi={10.1109/MM.2022.3140674}
}

@INPROCEEDINGS{Ramon_2021,
  title={{Ramon Space RC64-based AI/ML Inference Engine}},
  author={Ginosar, R. and Goldfeld, D. and Aviely, P. and Golan, R. and Meir, A. and Lange, F. and Alon, D. and Liran, T. and Shabtai, A.},
  booktitle={Eur. Workshop On-Board Data Process.},
  pages={1-33},
  year={2021},
  month={Jun},
  organization = {ESA-ESTEC},
  address = {Online Event}
}

@INPROCEEDINGS{Multi_Million_Core_2021,
  author={Lie, Sean},
  booktitle={IEEE Hot Chips Symp.}, 
  title={{Multi-Million Core, Multi-Wafer AI Cluster}}, 
  year={2021},
  month = {Aug.},
  organization={IEEE},
  address={Palo Alto, CA, USA},
  pages={1-41},
  doi={10.1109/HCS52781.2021.9567153}
}

@ARTICLE{Towards_Energy_Eff_Cluster_2012,
  author = {Lang, Willis and Harizopoulos, Stavros and Patel, Jignesh M. and Shah, Mehul A. and Tsirogiannis, Dimitris},
  title = {{Towards energy-efficient database cluster design}},
  year = {2012},
  volume = {5},
  number = {11},
  issn = {2150-8097},
  doi = {10.14778/2350229.2350280},
  journal = {ACM Proceedings of the VLDB Endow.},
  month = {July},
  pages = {1684–1695},
  numpages = {12}
}

@misc{nvidia_a100_2020,
  title = {{NVIDIA A100 Tensor Core GPU Architecture}},
  author = {{NVIDIA Corp.}},
  year = {2020},
  note = {[Online]. Available: \url{https://images.nvidia.com/aem-dam/en-zz/Solutions/data-center/nvidia-ampere-architecture-whitepaper.pdf}},
  howpublished = {NVIDIA Corp., Tech. Rep.}
}

@misc{nvidia_h100_2023,
  title = {{NVIDIA H100 Tensor Core GPU Architecture}},
  author = {{NVIDIA Corp.}},
  year = {2023},
  note = {[Online]. Available: \url{https://resources.nvidia.com/en-us-tensor-core}},
  howpublished = {NVIDIA Corp., Tech. Rep.}
}

@INPROCEEDINGS{interconnect_2020,
  author={Luan, Hao and Gatherer, Alan},
  booktitle={IEEE/ACM Int. Symp. Netw.-on-Chip}, 
  title={{Combinatorics and Geometry for the Many-ported, Distributed and Shared Memory Architecture}}, 
  year={2020},
  month={Nov.},
  organization={IEEE},
  address={Hamburg, Germany},
  pages={1-6},
  doi={10.1109/NOCS50636.2020.9241708}}

@ARTICLE{Mempool_2023,
  author={Riedel, Samuel and Cavalcante, Matheus and Andri, Renzo and Benini, Luca},
  journal={IEEE Trans. Comput.}, 
  title={{MemPool: A Scalable Manycore Architecture with a Low-Latency Shared L1 Memory}}, 
  year={2023},
  month={Aug.},
  volume={72},
  number={12},
  pages={3561-3575},
  doi={10.1109/TC.2023.3307796}
}

@INPROCEEDINGS{TeraPool_2024,
  author={Zhang, Yichao and Riedel, Samuel and Bertuletti, Marco and Vanelli-Coralli, Alessandro and Benini, Luca},
  booktitle={Great Lakes Symp. VLSI}, 
  title={{TeraPool-SDR: An 1.89TOPS 1024 RV-Cores 4MiB Shared-L1 Cluster for Next-Generation Open-Source Software-Defined Radios}}, 
  year={2024},
  month={June},
  pages={86-91},
  organization={ACM},
  address={Clearwater FL USA},
  doi={10.1145/3649476.3658735}
}

@ARTICLE{DRAMsys_2022,
author = {Steiner, Lukas and Jung, Matthias and Prado, Felipe S. and Bykov, Kirill and Wehn, Norbert},
title = {{DRAMSys4.0: An Open-Source Simulation Framework for In-depth DRAM Analyses}},
year = {2022},
volume = {50},
number = {2},
issn = {0885-7458},
doi = {10.1007/s10766-022-00727-4},
journal = {Int. J. Parallel Program.},
month = {April},
pages = {217-242},
numpages = {26}
}

@ARTICLE{Kung_1986,
  author =       {Kung, H. T.},
  title =        {{Memory Requirements for Balanced Computer Architectures}},
  year =         {1986},
  publisher =    {ACM},
  address =      {New York, NY, USA},
  volume =       {14},
  number =       {2},
  issn =         {0163-5964},
  doi =          {10.1145/17356.17362},
  journal =      {SIGARCH Comput. Archit. News},
  month =        {May},
  pages =        {49-54},
  numpages =     {6}
}

@ARTICLE{Scale_up_out_2016,
  author={Ma, Jun and Yan, Guihai and Han, Yinhe and Li, Xiaowei},
  journal={IEEE Trans. Comput.}, 
  title={{An Analytical Framework for Estimating Scale-Out and Scale-Up Power Efficiency of Heterogeneous Manycores}}, 
  year={2016},
  month={April},
  volume={65},
  number={2},
  pages={367-381},
  doi={10.1109/TC.2015.2419655}
}

@ARTICLE{tail_scale_2013,
author = {Dean, Jeffrey and Barroso, Luiz Andr\'{e}},
title = {{The Tail at Scale}},
year = {2013},
publisher = {ACM},
address = {New York, NY, USA},
volume = {56},
number = {2},
issn = {0001-0782},
doi = {10.1145/2408776.2408794},
journal = {Commun. ACM},
month = {Feb.},
pages = {74-80},
numpages = {7}
}

@ARTICLE{Illusion_2021,
  author = {R. M. Radway and A. Bartolo and P. C. Jolly and others},
  title = {{Illusion of Large On-Chip Memory by Networked Computing Chips for Neural Network Inference}},
  journal = {Nat. Electron.},
  volume = {4},
  number = {1},
  pages = {71--80},
  year = {2021},
  month={Jan.},
  doi = {10.1038/s41928-020-00515-3},
}

@ARTICLE{Occamy_2025,
  author={Scheffler, Paul and Benz, Thomas and Potocnik, Viviane and Fischer, Tim and Colagrande, Luca and Wistoff, Nils and Zhang, Yichao and Bertaccini, Luca and Ottavi, Gianmarco and Eggimann, Manuel and Cavalcante, Matheus and Paulin, Gianna and Gürkaynak, Frank K. and Rossi, Davide and Benini, Luca},
  journal={IEEE J. Solid-State Circuits}, 
  title={{Occamy: A 432-Core Dual-Chiplet Dual-HBM2E 768-DP-GFLOP/s RISC-V System for 8-to-64-bit Dense and Sparse Computing in 12-nm FinFET}}, 
  year={2025},
  month={Jan.},
  pages={1-15},
  note={Early access},
  doi={10.1109/JSSC.2025.3529249}
}

@ARTICLE{cop_2023,
  author={Li, Xin and Li, Zhi and Ju, Yaqi and Zhang, Xiaofei and Wang, Rongyao and Zhou, Wei},
  journal={IEEE Trans. Comput.}, 
  title={{COP: A Combinational Optimization Power Budgeting Method for Manycore Systems in Dark Silicon}}, 
  year={2023},
  month={Oct.},
  volume={72},
  number={5},
  pages={1356-1370},
  doi={10.1109/TC.2022.3211417}
}

@ARTICLE{Cdxbar_2019,
  author={Zhao, Xia and Ma, Sheng and Wang, Zhiying and Jerger, Natalie Enright and Eeckhout, Lieven},
  journal={IEEE Trans. Comput.}, 
  title={{CD-Xbar: A Converge-Diverge Crossbar Network for High-Performance GPUs}}, 
  year={2019},
  month={March},
  volume={68},
  number={9},
  pages={1283-1296},
  doi={10.1109/TC.2019.2906869}
}

@ARTICLE{Fpnew_2024,
  author={Bertaccini, Luca and Paulin, Gianna and Cavalcante, Matheus and Fischer, Tim and Mach, Stefan and Benini, Luca},
  journal={IEEE Trans. Emerg. Top. Comput.}, 
  title={{MiniFloats on RISC-V Cores: ISA Extensions with Mixed-Precision Short Dot Products}}, 
  year={2024},
  month={Feb.},
  volume={12},
  number={4},
  pages={1040-1055},
  doi={10.1109/TETC.2024.3365354}
}

@INPROCEEDINGS{block_matmul_2014,
  author={Ristov, Sasko and Gusev, Marjan and Velkoski, Goran},
  booktitle={Int. Conv. Inf. Commun. Technol. Electron. Microelectron.}, 
  title={{Optimal Block Size for Matrix Multiplication Using Blocking}}, 
  year={2014},
  month={July},
  organization={IEEE},
  address={Opatija, Croatia},
  pages={295-300},
  doi={10.1109/MIPRO.2014.6859580}
}

@ARTICLE{iDMA_2024,
author={Benz, Thomas and Rogenmoser, Michael and Scheffler, Paul and Riedel, Samuel and Ottaviano, Alessandro and Kurth, Andreas and Hoefler, Torsten and Benini, Luca},
journal={IEEE Trans. Comput.},
title={{A High-Performance, Energy-Efficient Modular DMA Engine Architecture}},
year={2024},
volume={73},
number={1},
ISSN={1557-9956},
pages={263-277},
doi={10.1109/TC.2023.3329930},
publisher={IEEE Computer Society},
month={Jan.}
}

@ARTICLE{manycore_edp_2018,
  author={Choi, Wonje and Duraisamy, Karthi and Kim, Ryan Gary and Doppa, Janardhan Rao and Pande, Partha Pratim and Marculescu, Diana and Marculescu, Radu},
  journal={IEEE Trans. Comput.}, 
  title={{On-Chip Communication Network for Efficient Training of Deep Convolutional Networks on Heterogeneous Manycore Systems}}, 
  year={2018},
  month={Nov.},
  volume={67},
  number={5},
  pages={672-686},
  doi={10.1109/TC.2017.2777863}
}

@ARTICLE{Programming_Model_2012,
  author={Diaz, Javier and Muñoz-Caro, Camelia and Niño, Alfonso},
  journal={IEEE Trans. Parallel Distrib. Syst.}, 
  title={{A Survey of Parallel Programming Models and Tools in the Multi and Many-Core Era}}, 
  year={2012},
  month={Jan.},
  volume={23},
  number={8},
  pages={1369-1386},
  doi={10.1109/TPDS.2011.308}
}

@INBOOK{Kalery_2021,
  author = {B. Dupont de Dinechin},
  title = {{A Qualitative Approach to Many-Core Architecture}},
  chapter = {2},
  pages = {27-51},
  editor = {L. Andrade and F. Rousseau},
  booktitle = {Multi-Processor System-on-Chip 1: Architectures},
  publisher = {Wiley},
  address = {Hoboken, New Jersey, USA},
  month = {April},
  year = {2021}
}

@INPROCEEDINGS{Ramon_2016,
  author={Ginosar, Ran and Aviely, Peleg and Israeli, Tsvika and Meirov, Henri},
  booktitle={IEEE Aerosp. Conf.}, 
  title={{RC64: High Performance Rad-Hard Manycore}}, 
  year={2016},
  month={March},
  organization={IEEE},
  address={Big Sky, MT, USA},
  pages={1-9},
  doi={10.1109/AERO.2016.7500697}
}

@ARTICLE{Tenstorrent_2021,
  author={Vasiljevic, Jasmina and Bajic, Ljubisa and Capalija, Davor and Sokorac, Stanislav and Ignjatovic, Dragoljub and et al., Thompson, David},
  journal={IEEE Micro},
  title={{Compute Substrate for Software 2.0}},
  year={2021},
  month={March},
  volume={41},
  number={2},
  pages={50-55},
  doi={10.1109/MM.2021.3061912}
}

@INPROCEEDINGS{nvidia_h100_uncovering_2024,
  author={Jin, Zhixian and Rocca, Christopher and Kim, Jiho and Kasan, Hans and Rhu, Minsoo and Bakhoda, Ali and Aamodt, Tor M. and Kim, John},
  booktitle={IEEE/ACM Int. Symp. Microarchitect.}, 
  title={{Uncovering Real GPU NoC Characteristics: Implications on Interconnect Architecture}}, 
  year={2024},
  month={Nov.},
  organization={IEEE},
  address={Austin, TX, USA},
  pages={885-898},
  doi={10.1109/MICRO61859.2024.00070}
}

@INPROCEEDINGS{blas_2016,
  author={Kepner, Jeremy and Aaltonen, Peter and Bader, David and Buluç, Aydin and Franchetti, Franz and Gilbert, John and Hutchison, Dylan and Kumar, Manoj and Lumsdaine, Andrew and Meyerhenke, Henning and McMillan, Scott and Yang, Carl and Owens, John D. and Zalewski, Marcin and Mattson, Timothy and Moreira, Jose},
  booktitle={HPEC}, 
  title={{Mathematical foundations of the GraphBLAS}}, 
  year={2016},
  month={Dec.},
  organization={IEEE},
  address={Waltham, MA, USA},
  volume={},
  number={},
  pages={1-9}
}

@INPROCEEDINGS{Hammerblade_2024,
  author={Jung, Dai Cheol and Ruttenberg, Max and Gao, Paul and Davidson, Scott and Petrisko, Daniel and et al., Taylor, Michael Bedford},
  booktitle={ISCA},
  title={{Scalable, Programmable and Dense: The HammerBlade Open-Source RISC-V Manycore}},
  year={2024},
  month={Aug.},
  organization={ACM/IEEE},
  address={Buenos Aires, Argentina},
  pages={770-784},
  doi={10.1109/ISCA59077.2024.00061}
}
